\documentclass{llncs}
\usepackage{latexsym,amssymb,enumerate,amsmath,epsfig,mathrsfs,stmaryrd}
\usepackage{subfigure}
\usepackage{tikz}
\usepackage{graphicx} 
\usepackage[ruled]{algorithm2e}
\usepackage{xcolor}
\usepackage{comment}

\newcommand{\abs}[1]{\left|#1\right|}
\newcommand{\env}{\mathcal{E}}
\newcommand{\pref}{{\sf Pref}}
\newcommand{\tl}{\text{tl}}

\newcommand{\Exp}{\mathbb{E}}
\newcommand{\ceil}[1]{\left\lceil#1\right\rceil}
\newcommand{\floor}[1]{\left\lfloor#1\right\rfloor}
\DeclareMathOperator*{\argmax}{argmax}

\newcommand{\dsum}{\displaystyle\sum}

\newcommand{\fq}{\text{fq}}

\newcommand{\VI}[1]{#1}

\title{Synthesis from LTL with Reward Optimization in Sampled Oblivious Environments\thanks{This work was supported by Fondation ULB, and the fund Thelam from the Fondation Roi Baudouin. Author Y.C. Tsai was previously funded by HKUST and the Lee Hysan Foundation, and is currently supported by the ERATO HASUO Metamathematics for Systems Design Project (No. JPMJER1603) and the ASPIRE grant No. JPMJAP2301, JST.}}
\author{}
\date{September 2024}

\author{Jean-Fran\c{c}ois Raskin$^1$ \and Yun Chen Tsai$^{2,3}$}

\institute{$^1$ Universit\'e libre de Bruxelles, Brussels, Belgium\\
$^2$ National Institute of Informatics, Tokyo, Japan\\
$^3$ The Graduate University for Advanced Studies (SOKENDAI), Hayama, Japan}

\begin{document}

\maketitle

\begin{abstract}
This paper addresses the synthesis of reactive systems that enforce hard constraints while optimizing for quality-based soft constraints. We build on recent advancements in combining reactive synthesis with example-based guidance to handle both types of constraints in stochastic, oblivious environments accessible only through sampling. Our approach constructs examples that satisfy LTL-based hard constraints while maximizing expected rewards—representing the soft constraints—on samples drawn from the environment. We formally define this synthesis problem, prove it to be {\sf NP-complete}, and propose an SMT-based solution, demonstrating its effectiveness with a case study.
\end{abstract}

\section{Introduction}

When designing reactive systems, we are often required to enforce {\em hard} constraints, such as those related to safety-critical features. In addition, we need to address more flexible, {\em softer} constraints, such as quality of service. While formal methods typically provide powerful tools for handling hard constraints, there are fewer solutions available for addressing soft constraints. In this paper, we build on recent work~\cite{DBLP:conf/tacas/BalachanderFR23} that demonstrates how to combine reactive synthesis, based on game theory and automata theory, with learning from examples. We extend this approach to propose a new tool for addressing {\em both} hard and soft constraints. We concentrate in this paper on the case of reactive synthesis with an unknown oblivious\footnote{An environment whose behavior is unaffected by the behavior of the reactive system.} and stochastic environment that is only accessible by sampling.

In~\cite{DBLP:conf/tacas/BalachanderFR23}, the following setting is studied. A linear temporal logic (LTL) formula, with atomic propositions partitioned into {\em inputs} controlled by an environment ({\sf Env}) and {\em outputs} controlled by a system ({\sf Sys}), defines the {\em core specifications} for a reactive system to be designed, where correctness must hold regardless of the environment's behavior. Alongside these core specifications, examples of desired system behaviors on sequences of environment inputs help guide the synthesis algorithm toward preferred solutions. This combination is advantageous because, while core correctness properties are easily expressible in LTL, specific implementation details, which may be challenging to capture in LTL, can be effectively illustrated through examples or scenarios. The algorithm proposed in~\cite{DBLP:conf/tacas/BalachanderFR23} addresses this synthesis problem by attempting to generalize the behaviors given in the examples, using automata learning techniques, while ensuring that these generalizations lead to a solution that enforce the core specification of the system.

In this paper, we consider another practically relevant design scenario. Here, we aim to synthesize a reactive system embedded in an oblivious environment, which needs to enforce a LTL formula (as above) which formalizes hard constraints for the system that needs to be enforced no matter what the environment does. However, instead of receiving a set of examples that demonstrate how to make "good" or "optimal" decisions on a set of prefixes of input sequences, we assume that we only receive a statistically representative set of samples of input sequences from the oblivious environment. Along with the samples, we are also given a reward machine ${\cal R}$, which maps prefixes of input/output sequences to rewards. The reward machine is used to formalize the {\em soft} constraints: the reactive system to be developed should aim to maximize the expected long-run average reward during its interaction with the environment while enforcing the hard constraints with certainty.

The goal of our procedure is to extend input samples from the environment into full prefixes of executions (alternating sequences of inputs and outputs) that serve as complete {\em examples} for the procedure in~\cite{DBLP:conf/tacas/BalachanderFR23}. The output choices used to complete these samples must meet two conditions: (1) the resulting set of examples must be extendable into a Mealy machine (a solution) that enforces the hard constraints specified by the LTL formula, so our completion of the input sequences must not compromise the realizability of the LTL specification; and (2) the expected mean reward across the samples is maximized. Once these examples are completed, we can apply the techniques from~\cite{DBLP:conf/tacas/BalachanderFR23}, which combine learning from examples and synthesis as recalled above, to generalize the decisions made on the samples and produce a full solution that enforces the LTL specification. Since these examples are completed with an emphasis on maximizing the expected reward, we anticipate that their generalization will lead to strong overall performance.

Our technical contributions are as follows. First, we formally define the problem described above. Second, we demonstrate that determining the optimal decisions for the set of input sequences obtained by sampling the environment is computationally intensive, even when the actions required to satisfy the hard constraints defined by the LTL specification are known. Specifically, we establish that this problem is {\sf NP-complete}. Third, we propose an SMT-based approach to solve the optimization problem. Finally, we illustrate our method with a case study.

\paragraph{{\bf  Related works }}
The LTL synthesis problem was first introduced in~\cite{DBLP:conf/popl/PnueliR89}. Since then, several works have contributed to the development of efficient algorithms to solve this problem, e.g.~\cite{DBLP:conf/focs/KupfermanV05,DBLP:conf/atva/ScheweF07a,DBLP:journals/fmsd/FiliotJR11,DBLP:journals/acta/LuttenbergerMS20}. Although the worst-case complexity is known to be high ({\sf 2ExpTime-complete}), tools for solving the LTL realizability problem have been implemented; see, for example, the tools described in the following papers: \cite{DBLP:journals/corr/abs-1803-09566,DBLP:conf/cav/BohyBFJR12,DBLP:journals/acta/LuttenbergerMS20,DBLP:conf/tacas/CadilhacP23}.

The algorithm presented in this paper builds on the new synthesis methods for LTL and example-based guidance introduced recently in~\cite{DBLP:conf/tacas/BalachanderFR23}. It demonstrates how to leverage these methods to synthesize a robust and efficient reactive system that interacts with an unknown, oblivious stochastic environment for which no model is available and only sampling is possible.

When a model of the stochastic environment is known, methods based on Markov Decision Processes (MDPs) can be applied to synthesize optimal reactive systems (see, e.g.,\cite{DBLP:books/daglib/0020348}). These methods have been implemented in tools such as {\sc Storm}\cite{DBLP:journals/sttt/HenselJKQV22} and {\sc Prism}~\cite{DBLP:journals/sigmetrics/KwiatkowskaNP09}.

MDP-based methods have been extended to handle multiple objectives, as in~\cite{DBLP:journals/acta/ChatterjeeRR14,DBLP:journals/fmsd/RandourRS17,DBLP:conf/tacas/QuatmannK21}. Additionally, the MDP model has been adapted to accommodate both properties that must be enforced with certainty (such as hard constraints) and those that cannot be guaranteed but for which we aim to maximize the probability of satisfaction (see, for example,~\cite{DBLP:conf/stacs/BruyereFRR14}). However, these approaches differ from those presented here, as they assume the existence of a model for the stochastic environment, whereas we assume access only to samples and not to a full model.

\paragraph{{\bf Structure of the paper}}
In Sect.~2, we introduce the necessary preliminaries for formally defining our setting and the problem we aim to solve. In Sect.~3, we present a formal problem definition along with a high-level overview of our algorithm. Our approach focuses on finding the optimal decisions for samples obtained from the stochastic environment, ensuring that these decisions also maintain the realizability of the hard constraints. In Sect.~4, we analyze the computational complexity of identifying these optimal decisions and propose a practical algorithm using SMT solvers. In Sect.~5, we demonstrate our approach with a case study.

\section{Preliminaries}

\subsection{LTL and the realizability problem}
We briefly review the syntax and semantics of LTL. For further details, interested readers may refer to, e.g.,~\cite{DBLP:books/daglib/0020348}. Given a set of atomic proposition ${\sf AP}$, a LTL formula over ${\sf AP}$ adheres to the following syntax:
$$\varphi:=p\in{\sf AP}~|~\neg \varphi~|~\varphi_{1}\lor \varphi_{2}~|~X \varphi~|~\varphi_{1} U \varphi_{2}$$
\noindent
Where $X$ is the next operator and $U$ is the until operator, we also define the globally operator $G(\varphi):=\neg ({\bf True}~U~\neg\varphi)$.

A LTL formula is evaluated over the positions on an infinite trace $\pi:=\pi_{1}\pi_{2}\pi_{3}... \in (2^{\sf AP})^{\omega}$, i.e. an infinite sequences of valuations, which defines for every index $i \in \mathbb{N}$, the set atomic propositions $\pi_i \subseteq {\sf AP}$ that hold in that index. 
The truth value of a formula $\varphi$ along a trace $\pi$ at index $i$ is defined as follows:
\begin{itemize}
    \item for $p \in {\sf AP}$, $\pi, i \models p \iff p\in \pi_{i}$
    \item $\pi, i\models X \varphi \iff \pi,i+1\models \varphi$
    \item $\pi, i\models \varphi_{1} U \varphi_{2}\iff \exists j\geq i \cdot (\pi,j\models \varphi_{2} \land \forall k\in[i,j) \cdot \pi, k\models \varphi_1)$
\end{itemize}
\noindent
Finally, a trace $\pi$ satisfies a LTL formula $\varphi$, denoted as $\pi\models \varphi$, if $\pi,1\models \varphi$, i.e. if $\pi$ satisfies $\varphi$ in its first position. Finally, we denote the set of traces that satisfies a formula $\varphi$ as follows:
$\llbracket\varphi\rrbracket := \{\pi:\mathbb{N}\mapsto 2^{\sf AP}|\pi\models \varphi\}$.

In the sequel, we partition the set of atomic propositions ${\sf AP}$ into ${\sf AP}_{I}\uplus{\sf AP}_{O}$, called {\em inputs} and {\em outputs} respectively.  Given ${\sf AP}_{I}$, we note $\Sigma_I$ the set of valuations for the atomic propositions in ${\sf AP}_{I}$, i.e. $\Sigma_I=2^{{\sf AP}_{I}}$, and  $\Sigma_O=2^{{\sf AP}_{O}}$, and we interprete a pair $(\sigma_I,\sigma_0) \in \Sigma_I \times \Sigma_O$ as a valuation for the entire set of atomic propositions ${\sf AP}$.

\paragraph{{\bf Realizability game}} 
Let $\psi$ be an LTL formula over the set of atomic propositions ${\sf AP} = {\sf AP}{I} \uplus {\sf AP}{O}$. The realizability game with objective $\psi$ is played by two players: {\sf Env}, who controls the atomic propositions in ${\sf AP}_{I}$ (also called inputs), and {\sf Sys}, who controls the atomic propositions in ${\sf AP}_{O}$ (also called outputs). The game proceeds for infinitely many rounds, and the interaction between the two players produces an infinite sequence $\pi$ of valuations over the atomic propositions ${\sf AP}$. Each valuation in this sequence is built over successive rounds. In each round $j \in \mathbb{N}$, {\sf Env} first chooses $\sigma_I \in \Sigma_I$, i.e., a valuation for the atomic propositions in ${\sf AP}_I$, and then {\sf Sys} responds by choosing $\sigma_O \in \Sigma_O$, a valuation for the atomic propositions in ${\sf AP}_O$. The pair $(\sigma_I, \sigma_O)$, which is a valuation for the entire set of atomic propositions in ${\sf AP}$, is appended to the sequence of valuations constructed in previous rounds. After infinitely many rounds, we thus obtain an infinite trace $\pi$ on which the truth value of the formula $\psi$ can be evaluated. {\sf Sys} wins the interaction if $\pi \models \psi$; otherwise, {\sf Env} wins. This game is therefore a zero-sum game.

The way {\sf Sys} plays in the game above can be formalized by the notion of a {\em strategy}, which is a function $\lambda_O: \Sigma_I^+ \rightarrow \Sigma_O$ that prescribes the choice $\lambda_O(h) \in \Sigma_O$ after a history of inputs $h = \sigma_{I,1}\sigma_{I,2} \dots \sigma_{I,n} \in \Sigma_I^+$, i.e., after a sequence of moves by Player {\sf Env}. Given an infinite sequence of valuations $\sigma_{I,1}\sigma_{I,2} \dots \sigma_{I,n} \dots \in \Sigma_I^{\omega}$ for the input variables in ${\sf AP}_I$, if {\sf Sys} plays according to $\lambda_O$, the outcome of the interaction is the following infinite sequence of valuations:
$$\pi=(\sigma_{I,1},\lambda_O(\sigma_{I,1})),
(\sigma_{I,2},\lambda_O(\sigma_{I,1},\sigma_{I,2})),\dots,
(\sigma_{I,n},\lambda_O(\sigma_{I,1}\sigma_{I,2}\dots\sigma_{I,n}))\dots$$
We note this outcome $\lambda_O(\sigma_{I,1}\sigma_{I,2}\dots\sigma_{I,n}\dots )$. In the sequel, we will say that given a prefix $(\sigma_{I,1},\sigma_{O,1})(\sigma_{I,2},\sigma_{O,2})\dots(\sigma_{I,n},\sigma_{O,n}) \in (\Sigma_I \times \Sigma_O)^*$ is compatible with strategy $\lambda_O$ if for all positions $i$, $1 \leq i \leq n$, $\sigma_{O,i}=\lambda_O(\sigma_{I,1} \sigma_{I,2} \dots \sigma_{I,i})$. 

We are now in position to recall the formal definition of the LTL realizabilty problem~\cite{DBLP:conf/popl/PnueliR89}. The {\em realizability} problem for a formula $\psi$ over ${\sf AP}={\sf AP}_{I} \uplus {\sf AP}_{O}$ asks if there exists a strategy $\lambda_O: \Sigma_I^+ \rightarrow \Sigma_O$ for ${\sf Sys}$ such that for all 
$\sigma_{I,1}\sigma_{I,2}\dots\sigma_{I,n}\dots \in (\Sigma_O)^{\omega}$, $$\lambda_O(\sigma_{I,1}\sigma_{I,2}\dots\sigma_{I,n}\dots) \models \psi$$ 
\noindent
(or equivalently if $\lambda_O(\sigma_{I,1}\sigma_{I,2}\dots\sigma_{I,n}\dots) \in \llbracket\psi\rrbracket$). In other words, the realizability problem asks whether there exists a strategy for {\sf Sys} such that all possible interactions with {\sf Env} result in an infinite trace that satisfies the specification. This problem is decidable as stated in the following theorem.

\begin{theorem}
    The realizability problem for LTL is {\sf 2ExpTime-C}. 
\end{theorem}

Like LTL, automata over infinite words define languages of infinite traces, or words, over a finite alphabet. From the definition of the realizability problem, it follows that we can replace the LTL formula with an $\omega$-regular language defined by an automaton over infinite words and evaluate the realizability of that language. Consequently, several automata-based solutions have been developed to decide the LTL realizability problem~\cite{DBLP:conf/popl/PnueliR89}. Here, we recall one approach based on {\em universal coB\"uchi automata}, variants of this approach were first introduced in~\cite{DBLP:conf/focs/KupfermanV05,DBLP:conf/atva/ScheweF07a,DBLP:conf/cav/FiliotJR09}.

\paragraph{{\bf Automata-based approach to realizablility}}
 Let $\Sigma=\Sigma_I \times \Sigma_O$, a {\em universal coB\"uchi automaton} over $\Sigma$ is a 5-tuple ${\cal A}=(Q,Q_{{\sf init}},\Delta,F)$ where: $Q$ is a finite set of states, $Q_{{\sf init}} \in 2^Q \setminus \emptyset$ is the non-empty set of initial states, $\Delta : Q \times \Sigma \rightarrow 2^Q \setminus \emptyset$ is the {\em universal} transition relation, and $F \subseteq Q$ is the set of B\"uchi states. A run of ${\cal A}$ over an infinite word $w=\sigma_1 \sigma_2 \cdots \sigma_n \dots \in \Sigma^{\omega}$ is an infinite sequence $r=q_1 q_2 \dots q_n \dots$ such that $q_1 \in Q_{{\sf init}}$, and for all indices $i \geq 1$, we have that $q_{i+1} \in \Delta(q_i,\sigma_i)$. According to the coB\"uchi acceptance condition, a run is {\em accepting} if the number of visits to $F$ along the run is {\em finite}, i.e. $\exists i \in \mathbb{N} \cdot \forall j \geq i : r(i) \not\in F$. A word $w$ is {\em accepted} by ${\cal A}$ if {\em all} its run on $w$ are accepting. In what follows, we also rely on a {\em stronger} acceptance condition, which is called the $K$-coB\"uchi condition, with $K \in \mathbb{N}_0$, which imposes that a run $r=q_1 q_2 \dots q_n \dots$ visits $F$ at most $K$ times (instead of finitely many times) to be accepting, i.e. $|\{ i \mid q_i \in F\}| \leq K$. We denote by $L({\cal A})$ the language defined by ${\cal A}$ under the coB\"uchi acceptance condition, and by $L({\cal A},K)$ the language defined by ${\cal A}$ with the $K$-coB\"uchi condition. Clearly, we have $L({\cal A},K) \subseteq L({\cal A})$ for all $K \in \mathbb{N}_0$, as the acceptance condition is strengthened.
For all LTL formula $\psi$ over ${\sf AP}$, we can construct an universal coB\"uchi automaton ${\cal A}_{\psi}$ such that $L({\cal A}_{\psi})=\llbracket\psi\rrbracket$. Thus, we can reduce the realizability of an LTL formula $\psi$ to the realizability of the language defined by the automaton ${\cal A}_{\psi}$. Furthermore, the following theorem extends this by linking the realizability of an LTL formula to the realizability of the associated \( K \)-coBüchi automata.

\begin{theorem}
Let $\psi$ be a LTL formula and ${\cal A}_{\psi}$ be a universal coB\"uchi automaton such that $L({\cal A}_{\psi})=\llbracket\psi\rrbracket$. Then for all $K \in \mathbb{N}_0$, if $\lambda_O$ realizes $L({\cal A}_{\psi},K)$ then $\lambda_O$ realizes $\psi$. Furthermore, if $\psi$ is realizable then there exists $K \in \mathbb{N}_0$, such that $L({\cal A}_{\psi},K)$ is realizable.
\end{theorem}

The advantage of the $K$-coB\"uchi approach lies in the fact that the automaton $({\cal A}_{\psi},K)$ defines a {\em safety language}. A word that does not belong to the language will have a finite prefix, and a partial run on this prefix that will exceed $K$ visits to $F$. As a consequence, the determinization of $K$-co-B\"uchi automata ${\cal A}$ relies on the following generalization of the subset construction: in addition to tracking the set of states that can be reached by a prefix of a run while reading an infinite word, the construction also counts the maximal number of times that a run prefix reaches states in the set $F$. Remember that a run can visit at most $K$ times such states to be accepting. The states of the deterministic automaton are called {\em counting functions} (that generalize subsets of states). These are formally defined for a co-B\"uchi automaton ${\cal A}=(Q,Q_{{\sf init}},\Sigma,\Delta,F)$ and $K \in \mathbb{N}_0$ as the set, denoted ${\sf CF}({\cal A},K)$, of functions $f : Q \rightarrow \{-1,0,1,\dots,K,K+1\}$. If $f(q)=-1$ for some state $q$, it indicates that $q$ is inactive (i.e., no run of ${\cal A}$ reaches $q$ on the current prefix). If $f(q)=x$, where $0 \leq x \leq K$, it means that $q$ is active, and the run leading to $q$ with maximal number of visits to $F$, has done $x$ visits to $F$. On the other hand, if $f(q)=K+1$, this indicates that a run on the current prefix has reached $q$ and visited $F$ more than $K$ times, implying that the word being read will be rejected.
The initial counting function $f_{{\sf init}}$ maps all initial states in $Q_{{\sf init}} \cap F$ to $1$, all other initial states to $0$, and all other states to $-1$ (which means that the state is not active). The deterministic automaton obtained by this determinization procedure is denoted by $\mathcal{D}({\cal A},K)=(Q^\mathcal{D} = CF({\cal A},K),q^\mathcal{D}_{{\sf init}},\Sigma,\Delta^\mathcal{D},Q^{\mathcal{D}}_{\sf reject})$, where the transition function $\Delta^\mathcal{D}$ follows the intuition developed above and $Q^{\mathcal{D}}_{\sf reject}$ are all the counting functions that maps a state $q$ to $K+1$. 

As the automaton $\mathcal{D}({\cal A},K)$ defines a safety language, its realizability problem is a {\em safety game}. During the realizability game, if by reading the prefixes of the word which is built by the interaction, the automaton never reaches $Q^{\mathcal{D}}_{\sf reject}$, {\sf Sys} wins, otherwise {\sf Env} wins. Solving this safety games, can be done with classical algorithms based on backward induction, as shown e.g. in~\cite{DBLP:journals/fmsd/FiliotJR11}. It is possible to extract from this analysis the subset of counting functions ${\sf Win} \subseteq {\sf CF}({\cal A},K)$ that are winning for {\sf Sys} in the associated safety game (all counting functions that are reached by prefixes of words from which {\sf Sys} has a winning strategy). Furthermore, if $f \in {\sf Win}$, then for all $f'$ such that, for all $q \in Q$, $f'(q) \leq f(q)$, then $f' \in {\sf Win}$, that is the set of winning counting functions is {\em downward closed}. This is a direct consequence of the fact that the language of suffixes accepted from $f'$ includes the set of prefixes accepted from $f$. Another interesting consequence of this fact is that we can equivalently represent ${\sf Win}$ by considering its maximal elements (as their downward closure is equal to ${\sf Win}$.) We write this set of maximal elements as $\ceil{\sf Win}$.
Details on how to compute symbolically the winning counting function can be found in~\cite{DBLP:journals/fmsd/FiliotJR11}.
As shown in that paper, the set of winning counting functions enjoys the following important properties.

\begin{theorem}\label{safeCF}
    Let $\psi$ be a LTL formula over ${\sf AP}_I \uplus {\sf AP}_O$, ${\cal A}_{\psi}$ be the coB\"uchi automaton over the alphabet $\Sigma=\Sigma_I \times \Sigma_O$ such that $L({\cal A})=\llbracket\psi\rrbracket$, $K \in \mathbb{N}_0$, and ${\sf Win}$ be the set of winning counting functions of the determinization of  $({\cal A}_{\psi},K)$, then if $h \in (\Sigma_I \times \Sigma_0)^*$ is such that by reading $h$ in $\mathcal{D}({\cal A}_{\psi},K)$, the run on $h$ stays wining ${\sf Win}$, then we know that there exists a strategy $\lambda_O$ for {\sf Sys} that realizes $\psi$ and is compatible with $h$.
\end{theorem}

\subsection{Pre-Mealy, Mealy and reward machines}

Strategies for the {\sf Sys} that uses finite memory can be encoded with Mealy machines. It is known that whenever a LTL specification is realizable, then it is realizable with a finite memory strategy and so with a Mealy machine.

A {\em pre-Mealy machine} $\mathcal{M}$ over $\Sigma_{I}$ and $\Sigma_{O}$ is defined as a $4$-tuple $\langle M,m_{0},\delta, {\cal O} \rangle$, where $M$ is a finite set of states, sometimes called {\em memory states}, $m_{0}\in M$ is the {\em initial state}, and $\delta:M\times\Sigma_{I}\mapsto M$, ${\cal O}:M\times \Sigma_{I}\mapsto \Sigma_{O}$ are partial functions called, respectively the {\em transition} function and the {\em output} function. A {\em Mealy machine} is a pre-Mealy machine with $\delta$ and ${\cal O}$ being complete, i.e. $\text{dom}(\delta)=\text{dom}({\cal O})=M\times\Sigma_{I}$. 

Let $\sigma_{I,1} \sigma_{I,2} \dots \sigma_{I,n} \in \Sigma_I^*$, the extended transition function $\delta^*$ of ${\cal M}$ is defined inductively as follows:
  \begin{itemize}
      \item if $n=1$, $\delta^*(m,\sigma_{I,1})=\delta(m,\sigma_{I,1})$
      \item if $n>1$, $\delta^*(m,\sigma_{I,1} \sigma_{I,2} \dots \sigma_{I,n})=\delta^*(\delta(m,\sigma_{I,1}),\sigma_{I,2} \dots \sigma_{i,n})$.
  \end{itemize}
We also define the finite language defined by a Mealy machine ${\cal M}$, noted $L^*({\cal M})$, and the infinite language, noted $L^{\omega}({\cal M})$, as follows:
  \begin{itemize}
      \item $L^*({\cal M})=\{ \sigma_{I,1} \sigma_{O,1} \dots \sigma_{I,n} \sigma_{O,_n} \in (\Sigma_I \times \Sigma_O)^* \mid \forall j \cdot 1 \leq j \leq n : \sigma_{O,j}={\cal O}(\delta^*(m_0,\sigma_{I,1} \dots \sigma_{I,j-1}),\sigma_{I,j})\}$
      \item $L^{\omega}({\cal M})=\{ \sigma_{I,1} \sigma_{O,1} \dots \sigma_{I,n} \sigma_{O,n} \dots \in (\Sigma_I \times \Sigma_O)^{\omega} \mid \forall j \in \mathbb{N} : \sigma_{O,j}={\cal O}(\delta^*(m_0,\sigma_{I,1} \dots \sigma_{I,j-1}),\sigma_{I,j})\}$
  \end{itemize}

A {\em reward machine} is a Mealy machine with input alphabet $\Sigma_{I} \times \Sigma_{O}$ that maps finite sequences from this alphabet to integer rewards. The output alphabet of the reward machine is defined by a pair of integers $r_{\min} < r_{\max}$, such that the output alphabet is the set $\mathbb{Z} \cap [r_{\min}, r_{\max}]$. We denote a reward machine by $\mathcal{R} = \langle S^{\mathcal{R}}, s_{{\sf init}}, \delta, r \rangle$, where $S^{\mathcal{R}}$ is a finite set of states, $s_{{\sf init}}$ is the initial state, $\delta : S^{\mathcal{R}} \times (\Sigma_{I} \times \Sigma_{O})) \rightarrow S^{\mathcal{R}}$ is the transition function, and $r : S^{\mathcal{R}} \rightarrow \mathbb{Z} \cap [r_{\min}, r_{\max}]$ is the reward function. For any finite word 
$\sigma_{I,1} \sigma_{O,1} \dots \sigma_{I,n} \sigma_{O,n} \in\left(\Sigma_{I}\times\Sigma_{O}\right)^{\ast}$, we define
\begin{equation}
\mathcal{R}\left(\sigma_{I,1} \sigma_{O,1} \dots \sigma_{I,n} \sigma_{O,n}\right):=\sum_{j=1}^{m}r\left(\delta^*(s_{{\sf init}},\sigma_{I,1} \sigma_{O,1} \dots \sigma_{I,n} \sigma_{O,j})\right)
\label{eq:reward}    
\end{equation}
to be the sum of rewards gained over the finite word $\sigma_{I,1} \sigma_{O,1} \dots \sigma_{I,n} \sigma_{O,n}$.

\subsection{Sample trees}
We assume that our environment, which is stochastic and oblivious, is modelled as a sequence of random variables $\{I_{i}\}_{i \in \mathbb{N}}$ that can be sampled. Samples are provided via a function $\env:\mathbb{N}\mapsto \Sigma_I^{*}$ which, for every $L\in\mathbb{N}$, returns a sequence $\env(L)$ of inputs of length $L$ from the environment: $\sigma_{I,1} \sigma_{I,2} \dots \sigma_{I,L} \in \Sigma_I^{L}$, so $\env(L)$ is itself a random variable that can be sampled multiple times. A sample obtained by sampling $\env(L)$ multiple times can be seen as a multiset\footnote{This is a multiset and not a set as identical sequences may appear multiple times when sampling $\env(L)$.} $S : \Sigma_I^L \rightarrow \mathbb{N}_0$ of finite sequences of input symbols of length $L$.

Let \( S \) be a sample obtained by drawing \( n \) observations from \( \env(L) \). We organize this sample, which consists of \( n \) input sequences of length \( L \), into a sample tree, defined as follows. Let \(\textsf{Pref}(S)\) denote the set of all prefixes of sequences that appear (possibly multiple times) in \( S \). The sample tree of \( S \), denoted \( T_S=(V,v_{\epsilon},E,p) \), is a tree with vertices \( V = \textsf{Pref}(S) \), representing the prefixes of input sequences found in \( S \). We denote the root of this tree by \( v_{\varepsilon} \) (the vertex associated with the empty word $\epsilon$).

The set of edges, \( E \subseteq V \times V \), contains an edge $(v, v') \in E$ if there exists \( \sigma_I \in \Sigma_I \) such that \( v' = v \cdot \sigma_I \). We refer to maximal prefixes \( v \in V \) as {\em leaves}, and a {\em branch} is defined as a path from the root to a leaf, we denote by ${\cal B}(T_S)$ the set of branches in $T_S$. Additionally, we define a function \( p : E \rightarrow [0,1] \) that assigns a probability to each edge, reflecting the relative frequency of prefixes in the sample \( S \), formally defined as follows:
$$  p(v,v')=\frac{\mbox{number of sequences in~} S \mbox{~with prefix~}v}{\mbox{number of sequences in~} S \mbox{~with prefix~}v'}
 $$
\noindent
Clearly, for all \( v \in V \), we have \(\sum_{v' \in V \mid (v,v') \in E} p(v, v') = 1\). Consequently, \( T_S \) forms a {\em finite Markov chain} with sink vertices, which correspond to the leaves of the tree.

\subsection{Partial strategy and expected reward}

Let \( T_S = (V, v_{\epsilon},E, p) \) be a sample tree for \( S \) that contains \( n \) input sequences of length \( L \) drawn from \( \env(L) \). A {\em partial \( \Sigma_{O} \)-strategy} over \( T_S \) is a function \( \sigma : V \setminus \{ v_{\varepsilon} \} \rightarrow \Sigma_{O} \). Intuitively, a partial strategy associates an output with each vertex of the sample tree, and thus with each prefix of inputs in \( \textsf{Pref}(S) \). Cleary, a partial strategy can be encoded as a Pre-Mealy machine.

Each branch \( b = v_{\varepsilon} v_1 v_2 \dots v_L \) in \( T_S \) corresponds to a sequence of inputs \( \sigma_{I,1} \sigma_{I,2} \dots \sigma_{I,L} \) in \( \textsf{Pref}(S) \). We denote by \( p(b) \) the probability of the branch \( b \) in $T_S$, calculated as \( \prod_{i=1}^{L-1} p(v_i, v_{i+1}) \).

Additionally, let \( \lambda_O(b) \) represent the sequence \( \sigma_{I,1} \sigma_{O,1} \sigma_{I,2} \sigma_{O,2} \dots \sigma_{I,L} \sigma_{O,L} \in (\Sigma_I \times \Sigma_O)^L \) where \( \sigma_{O,i} = \lambda_O(v_i) \) and \( v_i = \sigma_{I,1} \sigma_{I,2} \dots \sigma_{I,i} \), for all $i$, $1 \leq i < L$. Thus, \( \lambda_O(b) \) is the outcome of the strategy $\lambda_O$ on the sequences of inputs associated to the branch $b$.

Finally, let \( \mathcal{R}(\lambda_O(b)) \) denote the total sum of rewards obtained along branch \( b \) when following strategy \( \lambda_O \), as defined in equation~\ref{eq:reward}. We can now define the expected \( \mathcal{R} \)-reward of the partial strategy \( \lambda_O \) over \( T_S \) as follows:

\begin{equation}
    \Exp_{\lambda_O}(\mathcal{R},T_S):=\sum_{b\in{\cal B}(T_S)}p(b)\cdot\mathcal{R}(\lambda_O(b))\label{eqt:ER}
\end{equation}

\section{The problem and its algorithm}

We present the general problem we aim to solve, along with a high-level overview of the algorithm we use to address it. The problem is defined as follows:

\begin{problem} Given a realizable LTL formula $\psi$ over the set of atomic propositions ${\sf AP}={\sf AP}_I \uplus {\sf AP}_O$, a sample tree $T_S$ of input sequences sampled from the oblivious stochastic environment, and a reward machine ${\cal R}$, compute a strategy $\lambda_O$ that satisfies the following: 
$(i)$ realizes $\psi$, and $(ii)$ maximizes the expected total reward over the sample tree $T_S$.   
\end{problem}
\noindent
To obtain a solution to this problem, we proceed as follows:
\begin{itemize}
\item First, we compute the optimal outputs for each vertex in the sample tree to maximize the expected reward. This selection of optimal outputs defines a partial strategy that must also preserve the realizability of the LTL specification $\psi$, ensuring that the partial strategy can be extended into a complete strategy for ${\sf Sys}$ that satisfies $\psi$. We denote by ${\sf Ex} \subseteq (\Sigma_I \times \Sigma_O)^*$ the set of complete examples resulting from defining the partial strategy on $S$. Formally, ${\sf Ex} = \bigcup_{b \in {\cal B}(T_S)} \lambda_O(b)$.

\item Second, we apply the algorithm {\sf LearnSynth}, as defined in~\cite{DBLP:conf/tacas/BalachanderFR23}, to the set of examples ${\sf Ex}$ (obtained in the first phase) and the specification $\psi$ to derive a complete strategy that realizes $\psi$ and generalizes ${\sf Ex}$.
\end{itemize}

Let us make two important remarks here. First, as noted, we assume that $\psi$ is realizable; otherwise, there would be no solution to satisfy the hard constraints defined by $\psi$, and thus no solution to the overall problem we aim to solve. Additionally, we assume access to the universal coB\"uchi automaton ${\cal A}_{\psi}$, along with the maximal elements of a non-empty set of winning states $\lceil {\sf Win} \rceil$ for a given $K \in \mathbb{N}$. This additional input is a byproduct of testing the realizability of $\psi$ using a reactive synthesis tool such as {\sc Acacia}~\cite{DBLP:conf/cav/BohyBFJR12,DBLP:conf/tacas/CadilhacP23}, and hence incurs no additional cost in our context.

This setup allows us to apply Theorem~\ref{safeCF} to verify that the partial strategy defined on the sample tree can be extended into a complete strategy that realizes $\psi$. Specifically, a partial strategy is compatible with the set $\lceil {\sf Win} \rceil$ if the following condition holds: for all branches $b \in {\cal B}(T_S)$, when reading $\lambda_O(b)$ with the universal $K$-coB\"uchi automaton ${\cal A}_{\psi}$ associated with $\psi$, the corresponding counting functions remain within the downward closure of $\lceil {\sf Win} \rceil$.

Second, the purpose of using {\sf LearnSynth} to derive a complete strategy that realizes $\psi$ is to generalize the examples in {\sf Ex}. These examples demonstrate optimal decisions on input traces from the sample set among strategies that realize $\psi$. Since these examples represent optimal responses to inputs in the sample, applying {\sf LearnSynth} generalizes these decisions to produce a complete strategy—represented as a Mealy machine—which we expect to perform well in the unknown, oblivious environment.

To apply the above algorithm and use {\sf SynthLearn}, we must provide a method to optimally label a sample tree with the outputs of a partial strategy while ensuring the realizability of $\psi$. The worst-case complexity of this problem is analyzed in the following section, where we also present a practical solution to solve it.


\section{Computing the partial strategy}

In this section, we address the problem of computing a partial strategy for a sample tree \( T_S \) that meets two key requirements: \( (1) \) the partial strategy can be extended into a full strategy that realizes \( \psi \), meaning it maintains the realizability of \( \psi \); and \( (2) \) it maximizes the expected reward on \( T_S \) while satisfying the first requirement. The formal statement of this problem is given below:

\begin{problem}
 Given a realizable LTL formula $\psi$ over the set of atomic propositions ${\sf AP}={\sf AP}_I \uplus {\sf AP}_O$, a sample tree $T_S$ of input sequences sampled from the oblivious stochastic environment, and a reward machine ${\cal R}$,
 a unversal $K$coB\"uchi automaton ${\cal A}_{\psi}$ for the formula $\psi$, and a non-empty set of winning counting functions $\lceil {\sf Win} \rceil$ for $L({\cal A},K)$,
 compute a strategy $\lambda_O$ that satisfies the following: 
$(i)$ it maintains the realizability of $\psi$, and $(ii)$ maximizes the expected total reward over the sample tree $T_S$. 
\label{prob:realDeci}
\end{problem}

\subsection{Hardness}

We start by showing unless ${\sf PTime=NP}$, there is no polynomial time algorithm to solve problem~\ref{prob:realDeci}. Indeed, we prove in the next theorem that the decision version of this problem, which asks for the existence of a solution with an expected reward larger than or equal to a threshold $C \in \mathbb{Z}$ is {\sf NP-complete}.

\begin{theorem}The decision version of problem \ref{prob:realDeci} is {\sf NP-complete}.\label{thm:NPC}\end{theorem}
\noindent\textit{Proof}: The {\sf NP-membership} follows from the following arguments. Given a partial strategy $\lambda_{O}$, which has polynomial size relative to the sample tree and can therefore be guessed in polynomial time, we can verify two properties in deterministic polynomial time. First, to ensure that $\lambda_O$ preserves the realizability of $\psi$, we compute the counting functions reached by reading $\lambda_O(b)$ with ${\cal A}$ for all $b \in {\cal B}(T_S)$ and confirm that each of these counting functions falls within the downward closure of $\lceil {\sf Win} \rceil$. Second, we verify that the expected reward achieved by $\lambda_O$ on the Markov chain $T_S$ is at least $C$. This is accomplished by running the reward machine ${\cal R}$ on $\lambda_O(b)$ for each branch $b \in {\cal B}(T_S)$ to obtain its reward and then computing the weighted sum of these rewards based on the probability $p(b)$ of each branch $b$.

To establish {\sf NP-hardness}, we present a reduction from the graph independent set problem, which is defined as follows.


\begin{problem}[Graph Independent Set Problem] Given an undirected graph \( G = (U, H) \), where \( U \) is a finite set of vertices and \( H \) is a set of edges (pairs of adjacent vertices), and a parameter \( \kappa \in \mathbb{N} \), determine if there exists a subset of vertices \( I \subseteq U \) with cardinality at least \( \kappa \), i.e., \( \abs{I} \geq \kappa \), such that for all \( u, v \in I \), \( \{u, v\} \notin H \). \label{prob:gis} 
\end{problem}

Given a graph $G=(U,H)$ and a parameter $\kappa$, we now show how to reduce this instance to an instance of our decision problem. We start by fixing the input and output alphabet, we choose $\Sigma_{I}:=U=\{u_1 u_2 ... u_n\}$ and $\Sigma_{O}:=\{0,1\}$. 
We consider as the sample the unique sample sequence $u_1 u_2 \dots u_n$. Intuitively, a strategy $\lambda_O:V\setminus\{v_{\varepsilon}\}\mapsto \Sigma_{O}$ will assign $1$ after the history up to $u_{j}$ if the vertex $u_{j}$ is selected to be in the independent set $I$, and $0$ otherwise. We must then verify two properties. First, that $I$ is indeed an independent set in $G$. Second, that we have selected at least $\kappa$ vertices in $I$.
The first property will be coded in the universal $K$coB\"uchi automaton, while the second will be enforced by the reward machine. 

To ensure that the set of vertices selected by $\lambda_O$ forms an independent set, we encode the independence property using a $K$coB\"uchi automaton (with $K = 0$) over the alphabet $\Sigma_I \times \Sigma_O$. Additionally, we demonstrate that the winning counting functions for this universal $K$coB\"uchi automaton have a compact representation.

\begin{figure}[h!]
    \centering
    \subfigure[overview]{
        \begin{tikzpicture}[scale=0.13]
    \tikzstyle{every node}+=[inner sep=0pt]
    \draw [black] (21.1,-12.8) circle (3);
    \draw (21.1,-12.8) node {$q_{e_1}$};
    \draw [black] (21.1,-22) circle (3);
    \draw (21.1,-22) node {$q_{e_2}$};
    \draw (28,-35) node {$\vdots$};
    \draw [black] (21.1,-48.8) circle (3);
    \draw (21.1,-48.8) node {$q_{e_m}$};
    \draw [black] (34.8,-12.8) circle (3);
    \draw (34.8,-12.8) node {$q'_{e_1}$};
    \draw [black] (34.8,-22) circle (3);
    \draw (34.8,-22) node {$q'_{e_2}$};
    \draw [black] (34.8,-48.8) circle (3);
    \draw (34.8,-48.8) node {$q'_{e_m}$};
    \draw [black] (56.3,-31.3) circle (3);
    \draw (56.3,-31.3) node {$\perp$};
    \draw [black] (56.3,-31.3) circle (2.4);
    \draw [black] (58.817,-29.69) arc (150.34019:-137.65981:2.25);
    \fill [black] (59.11,-32.32) -- (59.56,-33.15) -- (60.05,-32.28);
    \draw [black] (24.1,-12.8) -- (31.8,-12.8);
    \fill [black] (31.8,-12.8) -- (31,-12.3) -- (31,-13.3);
    \draw [black] (37.07,-14.76) -- (54.03,-29.34);
    \fill [black] (54.03,-29.34) -- (53.75,-28.44) -- (53.09,-29.2);
    \draw [black] (24.1,-22) -- (31.8,-22);
    \fill [black] (31.8,-22) -- (31,-21.5) -- (31,-22.5);
    \draw [black] (37.55,-23.19) -- (53.55,-30.11);
    \fill [black] (53.55,-30.11) -- (53.01,-29.33) -- (52.61,-30.25);
    \draw [black] (24.1,-48.8) -- (31.8,-48.8);
    \fill [black] (31.8,-48.8) -- (31,-48.3) -- (31,-49.3);
    \draw [black] (37.13,-46.91) -- (53.97,-33.19);
    \fill [black] (53.97,-33.19) -- (53.04,-33.31) -- (53.67,-34.09);
    \draw [black] (9,-28.9) -- (19.3,-15.2);
    \fill [black] (19.3,-15.2) -- (18.42,-15.54) -- (19.22,-16.14);
    \draw [black] (9.69,-29.63) -- (18.61,-23.67);
    \fill [black] (18.61,-23.67) -- (17.66,-23.7) -- (18.22,-24.53);
    \draw [black] (9.07,-33.65) -- (19.23,-46.45);
    \fill [black] (19.23,-46.45) -- (19.13,-45.51) -- (18.35,-46.14);
    \end{tikzpicture}
        \label{fig:KCBA1}
    }%
    \subfigure[single component]{
        \begin{tikzpicture}[scale=0.2]
    \tikzstyle{every node}+=[inner sep=0pt]
    \draw [black] (20.6,-31.3) circle (3);
    \draw (20.6,-31.3) node {$q_{(v_i,v_j)}$};
    \draw [black] (31.8,-31.3) circle (3);
    \draw (31.8,-31.3) node {$q'_{(v_i,v_j)}$};
    \draw [black] (43.4,-31.3) circle (3);
    \draw (43.4,-31.3) node {$\perp$};
    \draw [black] (43.4,-31.3) circle (2.4);
    \draw [black] (21.923,-33.98) arc (54:-234:2.25);
    \fill [black] (19.28,-33.98) -- (18.4,-34.33) -- (19.21,-34.92);
    \draw [black] (23.6,-31.3) -- (28.8,-31.3);
    \fill [black] (28.8,-31.3) -- (28,-30.8) -- (28,-31.8);
    \draw (26.2,-30.8) node [above] {$v_i$};
    \draw [black] (34.8,-31.3) -- (40.4,-31.3);
    \fill [black] (40.4,-31.3) -- (39.6,-30.8) -- (39.6,-31.8);
    \draw (37.6,-31.8) node [below] {$v_j$};
    \draw [black] (33.123,-33.98) arc (54:-234:2.25);
    \fill [black] (30.48,-33.98) -- (29.6,-34.33) -- (30.41,-34.92);
    \draw [black] (46.08,-29.977) arc (144:-144:2.25);
    \fill [black] (46.08,-32.62) -- (46.43,-33.5) -- (47.02,-32.69);
\end{tikzpicture}
        \label{fig:KCBA2}
    }
    \caption{illustration of the automata}
\end{figure}
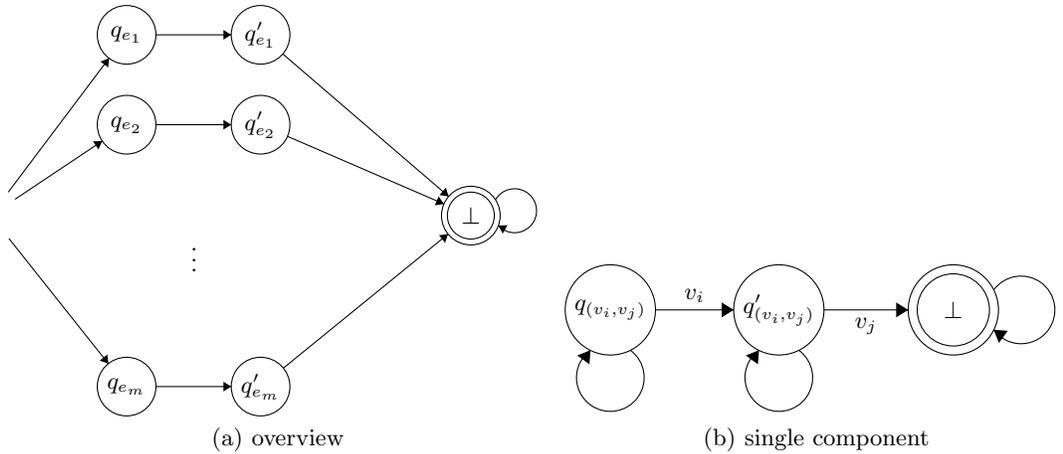\noindent

First, we formally define the automaton,
\begin{definition}
    Given a undirected graph $G=(U,H)$, we can construct an automaton $\mathcal{A}:=(Q,q_{\sf init},F,\delta^{\mathcal{A}})$ for checking whether a string $(U\cdot\{0,1\})^{*}$ encodes an independent set $I\subset U$. The construction is defined as follow, $Q:=\{q_e,q_{e}'|e\in F\}\cup\{\perp\}$, $q_{\sf init}=\{q_e|e\in F\}$, $F=\{\perp\}$ and $\delta^{\mathcal{A}}$ as,
    \begin{align*}
      \delta^{\mathcal{A}}(q_{e}, \ast 0) &= \{q_{e}\}\\
      \delta^{\mathcal{A}}(q_{(v_i,v_j)}, v_t 1) &= \begin{cases}
          \{q_{(v_i,v_j)}'\}&\quad t=i\\
          \{q_{(v_i,v_j)}\}&\quad t\neq i
      \end{cases}\\
      \delta^{\mathcal{A}}(q_{e}', \ast 0) &= \{q_{e}\}\\
      \delta^{\mathcal{A}}(q_{(v_i,v_j)}', v_t 1) &= \begin{cases}
          \{\perp\}&\quad t=j\\
          \{q_{(v_i,v_j)}'\}&\quad t\neq j
      \end{cases}\\
      \delta^{\mathcal{A}}(\perp,\ast\ast) &= \{\perp\}
    \end{align*}
    where $\ast$ represent arbitrary letter in the alphabet, and we assumed that for any $e=(v_{i},v_{j})$, we have $i<j$.\label{def:GISauto}
\end{definition}

For each edge $\{u_i, u_j\} \in H$ with \( i < j \), the universal \( 0 \)-coB\"uchi automaton has an initial state \( q_{ij} \) where control remains (with appropriate self-loops) until the letter \((u_i, 1)\) is read, indicating that vertex \( u_i \) has been selected for the independent set. Upon reading this letter, the automaton transitions to a state where it waits for the letter \((u_j, 1)\) (with additional self-loops). If this second letter is read, signifying that vertex \( u_j \) has also been selected, the automaton moves to a state in $F$, where it remains indefinitely, thus violating the \( 0 \)-coB\"uchi condition. Clearly, if such a state in $F$ is reached, it means that the strategy $\lambda_O$ has selected two vertices connected by an edge in set \( I \), which makes the choice invalid. This setup forces the universal automaton to reject any continuation of $\lambda_O(u_1 u_2 \dots u_n)$, preventing us from completing the partial strategy $\lambda_O$ into a strategy that realizes the language defined by the universal automaton.

Conversely, if the local strategy $\lambda_O$ selects vertices that form an independent set, no run reaches a state in $F$ upon completing the sample, allowing $\lambda_O$ to be trivially extended into a strategy that realizes the language defined by the \( K \) universal automaton, as no accepting state is ever reached. The automaton is illusrated in figure (\ref{fig:KCBA1}) and (\ref{fig:KCBA2}).

From this description of the universal automaton, it follows that all counting functions where no active state in $F$ are winning, as {\sf Sys} can always emit \( 0 \) (indicating it is not selecting any more vertices). There is a unique maximal counting function that represents this set of winning functions: all states $q$ not in $F$ are mapped to \( 0 \), while the states in $F$ are mapped to \( -1 \) (indicating they are inactive). Formally we prove the following lemma,

\begin{lemma}
    Given the automata constructed in definition \ref{def:GISauto}, its upward maximal antichain of realizable counting function $\ceil{W_{0}^{\mathcal{A}}}$ is $\{f^{*}\}$ where
    $$f^{*}(q)=\begin{cases}
        -1&\quad q=\perp\\
        0&\quad \text{otherwise}
    \end{cases}$$\label{lem:RealAnti}
\end{lemma}

\textit{Proof}: From the definition of the automaton, as all the state loops in itself upon a $0$ output is received, hence it is trivial to construct a strategy to win the safety game from $f^*$, namely $\lambda_O(u\in U) = 0\quad\forall u\in U$, hence we have $f^*\in W_{0}^{\mathcal{A}}$. Next we need to show that $\forall f\in {\sf Win}$, $f\preceq f^*$. On the contrary, assume $\exists f$ such that $f\not\preceq f^*$. By definition that $f \in {\sf Win}\implies f(q)\leq K=0\quad\forall q$, thus for all $q\neq \perp$, we have $f(q)\leq 0 = f^{*}(q)$. Hence in order to have $f(q)\not\preceq f^*$, it forces that $-1=f^*(\perp)< f(\perp) \leq 0\implies f(\perp) = 0$. Pick any strategy $\lambda_O$ and input $u\in U$, let $f' = \delta^{\mathcal{D(A)}}(f,u\sigma(u))$, where $\delta^{\mathcal{D(A)}}$ denote the transition function in the determinized automaton with counting function as state. By definition of the automaton, the state $\perp$ always loop in itself regardless of the input, hence we must have $f'(\perp) \geq f(\perp) + \chi_{Q_usf}(\perp) = 1$, which immediately implies that $f'\notin {\sf Win}$. Since this holds for arbitrary strategy $\lambda_O$, thus there is no strategy that realize $\mathcal{A}$ from $f$, which further implies that $f\notin {\sf Win}$ and thus contradicting our assumption. \qed

We still need to encode the requirement that $\lambda_O$ must select at least \(\kappa\) vertices to form an independent set with the appropriate size. This can be accomplished by defining a reward machine \({\cal R}\) that rewards $\lambda_O$ each time it emits the letter \(1\), indicating a vertex selection. By setting a reward threshold \(C = \kappa\), we ensure that $\lambda_O$ selects at least \(\kappa\) vertices.

With these constructions, we are now ready to prove the equivalence of solution between the two instance of problems. Considering having an independent set $I$ with $\abs{I}\geq \kappa$, then we construct a strategy $\lambda_O(v_i)=\chi_{I}(v_i)$ where the $\chi_I$ represents the characteristic function of $I$ which gives $1$ if $v_i\in I$ and $0$ otherwise. Let we note that it gives $1$ for all $v_i\in I$ thus one can easily see that the corresponding string gives exactly a reward of $\kappa$. Further it is safe with respect to the automata, since $\forall u,v\in E$, $(u,v)\notin E$ and in order to reach $\perp$ in one of run of the automata, the string must contains $u1$ and $v1$ for some $(u,v)\in E$ which cannot happen as we only gives the output $1$ when $u\in I$, thus the counting function reached after the finite run must be lower than the $f^*$ defined above and thus being realizable by lemma \ref{lem:RealAnti}. This completed part of the proof.

For the other direction, consider having a realizable strategy $\lambda_O$ such that $\Exp_{\lambda_O}(\mathcal{R},T_{S})\geq k$. Then we define the set $I:=\{v\in V|\lambda_O(v)=1\}$, by the definition of the reward machine, we must have $\abs{I}\geq k$, hence it left for us to verify that $I$ is an independent set. Assuming it is not, then $\exists v_i,v_j\in I$ such that $(v_i,v_j)\in E$, without loss of generosity assume that $i < j$, then upon the run that start at node $q_{v_i,v_j}$, it must move to state $q_{v_i,v_j}'$ at the point reading $v_i1$ in the sample input since $v_i\in I\iff \lambda_O(v_i) = 1$, with the same reasoning the automata will eventually reach state $\perp$ and hence losing the safety game, which contradicts the assumption that $\lambda_O$ being realizable, hence finishing our proof. \qed

~\\
Therefore, we cannot expect to solve Problem~\ref{prob:realDeci} with a polynomial-time algorithm unless {\sf PTime=NP}. Let us note that this result is somewhat surprising, as it can be shown that verifying the existence of a local strategy that meets condition $(1)$ can be done efficiently (since we have access to the universal automaton $\mathcal{A}_{\psi}$ and the set $\lceil {\sf Win} \rceil$). Similarly, computing an optimal local strategy that maximizes the expected reward on \( T_S \) can also be done in polynomial time (by treating \( T_S \) as a Markov Decision Process (MDP)). Thus, it is the combination of these two constraints that makes the problem {\sf NP-complete}.

\subsection{SMT based solution}
Now that we have established that Problem~\ref{prob:realDeci} is {\sf NP-complete}, we show how to solve it using a SMT solver.

We start by providing some intuition. To encode the decision version of Problem~\ref{prob:realDeci}, we need to track the execution of two machines along the branches of $T_S$, which are annotated by the choices made by the partial strategy $\lambda_O$: the universal automaton, to monitor the counting functions reached, and the reward machine, to compute the reward. However, since the partial strategy is not yet fixed, and discovering is the purpose of the SMT based procedure, we maintain two sets of variables, \VI{one for representing the output choices on each vertex of the sample tree and one for tracking the counting function. The constraint should specify rules for setting the values of the variable so that the transition on all machines are correctly simulated and the realizability are ensured.}

Formally, given a universal $K$-coB"uchi automaton $\mathcal{A}$ with access to the antichain of realizable counting functions $\ceil{\sf Win}$, a sample tree $T_S$, and a reward machine ${\cal R}$, we define two sets of variables. The first set consists of boolean variables used to determine the strategy to be applied at each vertex $v$:
$$\{x_{v,s,o}\mid \forall v\in V, S\in S^{\cal R}, o\in \Sigma_{O}\}$$
The second set consists of integer variables used to track the counting function after executing the action determined by the previous set of variables:
$$\{y_{v,q}\mid \forall v\in V, q\in Q\}$$
We then encode the initial conditions, transitions, and realizability constraints and pass these to an SMT solver to obtain a partial strategy $\lambda_{O}$ that solves the decision version of Problem~\ref{prob:realDeci} with respect to a given parameter $C$. Now we proceed to outline the encoding,

\subsubsection*{Ambiguity}
First, to ensure that exactly one action is chosen for each $v\in V_{I}\setminus\{v_\varepsilon\}$, $\forall v\in V_{I}\setminus\{v_\varepsilon\}$, the ambiguity constraint is formulated as
\begin{equation}
    \sum_{s\in S}\sum_{o\in O}x_{v,s,o}=1\label{eqt:amb}
\end{equation}
it's not hard to see that the equality holds if and only if exactly one of the variables among $\{x_{v,s,o}\}$ is chosen for each vertex $v$.
\subsubsection*{Transition correctness in reward machine}
Next we handle the transition in the reward machine, we start by ensuring the initial state is correctly set for all vertices that are one step under the root, namely $\forall v\in \{v|(v_{\varepsilon},v)\in E\}$,,
\begin{equation}
    \sum_{o\in\Sigma_{O}} x_{v,s_{0},o}= 1\label{eqt:tr0}
\end{equation}

Next we handle the general transition, consider that if a certain output $o$ is fixed at a vertex $v$ in the sample tree associate with some state $s$ of the reward machine, then it deterministically decided the state in the reward machine for all its successors on the sample tree. Hence symbolically it can be formulated as $\forall v\in V$, $\forall s\in S^{\cal R}, o\in\Sigma_{O}$, let $s'=\delta^{\mathcal{R}}(s,\tl(v)o)$ which is the successor state, $\forall u\in \alpha^{-1}(v)$,
\begin{equation}
    x_{v,s,o}=1\implies \sum_{o'\in\Sigma_{O}}x_{u,s',o'}=1\label{eqt:tr1}
\end{equation}
Then we can first dropped the $=1$ and regards the variables here as boolean variable, for the summation noted that by equation (\ref{eqt:amb}) it cannot exceed $1$ thus we can safely perform the transformation. Then we have the equivalent form of $x_{v,s,o}\implies \sum_{o'\in\Sigma_{O}}x_{u,s',o'}$, and we noted that any logical statement of the form $p\implies q$ can be equivalently expressed as a numerical inequality of the form $(1-p)+q\geq 1$, hence we have
\begin{equation}
    (1-x_{v,s,o})+\sum_{o'\in\Sigma_{O}}x_{u,s',o'}\geq 1\label{eqt:tr2}
\end{equation}
\subsubsection*{Realizability}
Now we procceed to handle the realizability, for that we manipulate the set of variables $\{y_{v,q}\}$ to encode the counting function. As similar to the encoding of transition in reward machine, we have to consider various issues, namely i) The initial counting is set correctly, ii) the transition is correctly modeled, iii) no unrealizable counting function is visited, we will address them one by one in the following.

\noindent
The initial condition is relatively simple, we simply require the encoding at $v_{\varepsilon}$ correctly corresponds to the initial counting function, this can be encoded as $\forall q\in Q$
\begin{equation}
    y_{v_{\varepsilon},q}=\begin{cases}
        -1&\quad q\notin q_{\sf init}\\
        1&\quad q\in q_{\sf init}\cap F\\
        0&\quad \text{otherwise}
    \end{cases}\label{eqt:autoInit}
\end{equation}
Noted that the conditioning here is for brevity of presentation and one shall see that the conditions are mutually exclusive thus in practice these will be simple assertion of the form $y_{v_{\varepsilon},q}=c$.

The case for general transition is slightly more complicated as we need to relate the counting function between subsequent state. Fix any non-root vertex $v\in V\setminus\{v_{\varepsilon}\}$ on the sample tree and let $u$ be the unique ancestor of $v$, i.e. $(u,v)\in E$. Consider an output $o$ is chosen for the vertex $v$, then for each state $q\in Q$ in the original co-B\"uchi automaton, the new counting function checks whether $q$ is reachable from the counting function at $u$ and update it according to the definition of the counting function. Thus we start by consider the encoding of the reachability from active state, denote $i_v\in\Sigma_{I}$ as the last letter of the prefix on $v$, then
\begin{equation}
    \sum_{p : q\in\Delta(p,i_v o)}(y_{u,p}+1)>0\label{eqt:autoReach}
\end{equation}
encodes the reachability of $q$ from the counting function on $u$. The idea behind is as follow, the state $q$ is reachable by some active state if and only if one of it ancestor, captured by the set $\{p\in Q|q\in\Delta(p,i_v o)\}$, contains at least one active state. Recall from the definition of counting function, a state $p$ is active in a counting function $f$ if and only if $f(p)>-1$, which is equivalent to $y_{u,p}>-1 \equiv y_{u,p}+1 >0$ in our variable encoding. Since we only require one active state to have the reachability, thus summing over all ancestor state $p$ decide whether the state $q$ is reachable from the counting function encoded by $\{y_{u,p}\}_{p\in Q}$. Applying this, we construct the following two constraints. $\forall v\in V\setminus\{v_{\varepsilon}\}$, $\forall s\in S^{\cal R}, o\in\Sigma_{O}$, $\forall u\in \{u\in V|(u,v)\in E\}, q\in Q$,
\begin{equation}
    x_{v,s,o}=1\implies \left(\dsum_{p:q\in\Delta(p,i_v o)}(y_{u,p}+1)=0\iff y_{v,q}=-1\right)\label{eqt:autoTr1}
\end{equation}
and $\forall p\in \{p\in Q|q\in\Delta(p,i_v o)\}$,
\begin{equation}
    x_{v,s,o}=1\implies \left(y_{u,p}\geq 0\implies y_{v,q}\geq y_{u,p}+\chi_{usf}(q)\right)\label{eqt:autoTr2}
\end{equation}
Noted that here we dropped the case wher $y_{v,q}$ takes value $K+1$, since we are searching for solution that is realizable, thus having the value $K+1$ will automatically violate the constraint for checking realizable, which shall be formulate below, $\forall v\in V$
\begin{equation}
    \bigvee_{f\in \ceil{\sf Win}}\bigwedge_{q\in Q} y_{v,q}\leq f(q)\label{eqt:autoReal}
\end{equation}

\noindent
The above constraints (\ref{eqt:amb}), (\ref{eqt:tr0}), (\ref{eqt:tr2}), (\ref{eqt:autoInit}), (\ref{eqt:autoTr1}), (\ref{eqt:autoTr2}) and (\ref{eqt:autoReal}) ensured that the variables encode a valid and realizable strategy and we are now left to account for the optimality. Since we intend to construct a SMT instance, we do not solve the optimization problem directly. Instead we consider the decision variant as defined in problem \ref{prob:realDeci}, then we add the last constraint on the expected reward of the strategy. By induction, one can show the following
\begin{equation}
    \Exp_{\sigma}(\mathcal{R},T_{n,L})=\sum_{v\in V,s\in S, o\in\Sigma_{O}}r(v,s,o)\cdot\frac{\fq(v)}{n}\cdot x_{v,s,o}\label{eqt:TER2SMT}
\end{equation}
where $\fq(v)$ is defined as the number of sequences in $S$ with prefix $v$. Then we rearrange it and construct the following constraint
\begin{equation}
    \sum_{v\in V,s\in S, o\in\Sigma_{O}}r(v,s,o)\cdot\fq(v)\cdot x_{v,s,o}\geq nC\label{eqt:optimal}
\end{equation}
This completes our reduction to SMT, it is not hard to see that the reduction is indeed poloynomial in the input size as there are only polynomial number of constraints. Assuming there is only a small constant number of realizable counting function in the antichain $\ceil{\sf Win}$, the time for constructing all the constraints will be dominated by the time needed for equation (\ref{eqt:autoTr2}), which is at most $\abs{S^{\cal R}}\abs{\Sigma_{O}}\abs{\{u\in V|(u,v)\in E\}}\abs{Q}^{2}$ for each $v\in V\setminus\{v_{\varepsilon}\}$. Noted that each $\{u\in V|(v,u)\in E\}$ corresponds to an edge in $E$ and there is clearly no double counting since it is a tree, hence summing all up gives $O(\abs{S^{\cal R}}\abs{\Sigma_{O}}\abs{V}\abs{Q}^{2})$ time.

\subsection{Solving the Optimality}
Recall that problem \ref{prob:real} is originally an optimization problem, to address this we start by noting that the number of transition in the reward machine $R$ is bounded, with $r_{\min}$ and $r_{\max}$ as the minimum and maximum reward among all the transitions, then the left hand side of the above inequality is trivially bounded between the following,
$$r_{\min}nL\leq \sum_{v\in V,s\in S^{\cal R}, o\in\Sigma_{O}}r(v,s,o)\cdot\fq(v)\cdot x_{v,s,o}\leq r_{\max}nL$$
Hence by adjusting the value of $C$, one can perform binary search over the interval $[r_{\min}nL,r_{\max}nL]$ to obtain the optimal solution, the expression after transform is of integer value thus it takes $O(\log((r_{\max}-r_{\min})nL))$.

\section{Case study: weather monitoring system}

In this section, we illustrate how our formal framework and synthesis algorithm apply to a system that must meet {\em both} hard and soft constraints while interacting with an unknown, oblivious stochastic environment.

The design scenario is as follows: the objective is to automatically synthesize a weather monitoring system capable of issuing {\em warnings} and {\em alarms} to alert road administration teams of freezing hazards, enabling timely deployment of salt trucks and other preventive measures when road conditions become unsafe. In this scenario, {\sf Env} controls the evolution of temperature, while the {\sf Sys} manages the warning and alarm signals. Clearly, the {\sf Env} is {\em oblivious} to the behavior of {\sf Sys}, as it is independent of the system’s decisions; that is, temperature evolution is unaffected by the issuance of alarms or warnings.

Similar to real-world conditions, the monitoring system is developed without a fully defined probabilistic model of temperature evolution. Instead, the system relies on previous observations of temperature patterns, leveraging statistical trends to optimize warning issuance based on historical data. This historical data will be played by our sample tree.

According to our framework, the specifications governing the system’s operation are divided into two categories: {\em hard constraints} related to {\em alarms} and {\em soft constraints} related to {\em warnings}.

\paragraph{Informal description}
We begin with a description of the hard constraints. An {\em alarm} is triggered whenever the temperature reaches or falls below 0 degrees Celsius. Once active, the alarm remains on until the temperature rises above 0 degrees Celsius for at least two consecutive time steps, after which it automatically deactivates. While an alarm is active, all warnings are disabled to prevent redundancy. Below, we show how to formalize these hard constraints in LTL.

Now, let us turn to the {\em soft constraints} related to warnings. A {\em warning} is triggered when the temperature is nearing 0 degrees but has not yet reached it. A warning is considered valid {\em in insight} if the temperature subsequently reaches 0 degrees; otherwise, it is deemed a false alarm if the temperature rises again without reaching the critical threshold. To assess system performance, a cost function evaluates the accuracy of warnings, penalizing false alarms (negative rewards) while incurring no penalty for legitimate warnings.

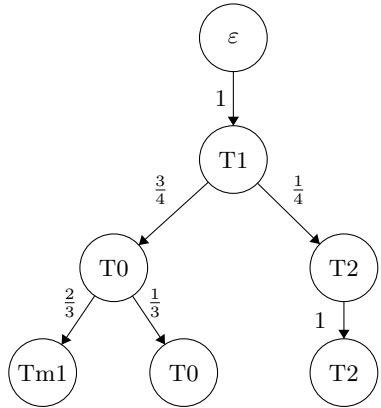
\begin{figure}[h!]
    \centering
    \begin{tikzpicture}[scale=0.15]
    \tikzstyle{every node}+=[inner sep=0pt]
    \draw [black] (38.7,-6) circle (3);
    \draw (38.7,-6) node {$\varepsilon$};
    \draw [black] (28.1,-26.4) circle (3);
    \draw (28.1,-26.4) node {T0};
    \draw [black] (21.8,-35.7) circle (3);
    \draw (21.8,-35.7) node {Tm1};
    \draw [black] (34.3,-35.7) circle (3);
    \draw (34.3,-35.7) node {T0};
    \draw [black] (48.5,-26.4) circle (3);
    \draw (48.5,-26.4) node {T2};
    \draw [black] (48.5,-35.7) circle (3);
    \draw (48.5,-35.7) node {T2};
    \draw [black] (38.7,-16.8) circle (3);
    \draw (38.7,-16.8) node {T1};

    \draw [black] (38.7,-9) -- (38.7,-13.8);
    \fill [black] (38.7,-13.8) -- (39.2,-13) -- (38.2,-13);
    \draw (38.2,-11.4) node [left] {$1$};

    \draw [black] (36.48,-18.81) -- (30.32,-24.39);
    \fill [black] (30.32,-24.39) -- (31.25,-24.22) -- (30.58,-23.48);
    \draw (33,-19.5) node [left] {$\frac{3}{4}$};

    \draw [black] (40.84,-18.9) -- (46.36,-24.3);
    \fill [black] (46.36,-24.3) -- (46.14,-23.38) -- (45.44,-24.1);
    \draw (43.7,-19.5) node [right] {$\frac{1}{4}$};

    \draw [black] (26.42,-28.88) -- (23.48,-33.22);
    \fill [black] (23.48,-33.22) -- (24.35,-32.83) -- (23.52,-32.27);
    \draw (24.95,-29.5) node [left] {$\frac{2}{3}$};
    
    \draw [black] (29.76,-28.9) -- (32.64,-33.2);
    \fill [black] (32.64,-33.2) -- (32.61,-32.26) -- (31.78,-32.82);
    \draw (31,-29.5) node [right] {$\frac{1}{3}$};
    
    \draw [black] (48.5,-29.4) -- (48.5,-32.7);
    \fill [black] (48.5,-32.7) -- (49,-31.9) -- (48,-31.9);
    \draw (47,-31.05) node [left] {$1$};
\end{tikzpicture}
    \label{fig:SampT}
    \caption{An example of sample tree.}
\end{figure}

While it is straightforward to design a system that meets the hard constraints for alarms only, achieving optimal adherence to the soft constraints for warnings is more complex. Consider the following multiset of samples (eight sequences of four inputs each) obtained from the environment: 
$$S(2;1;0;-1)=2,S(2;1;0;0)=1,S(2;1;2;2)=1,S(2;2;2;2)=3,S(2;2;1;2)=1$$
which correspond to the sample tree depicted in Fig.\ref{fig:SampT}. In the node labeled $1$ on the left branch, there is a higher likelihood that the temperature will reach 0 degrees, so the system may issue a warning. However, there is also a chance that the temperature will subsequently rise to $2$, rendering the warning unnecessary and suboptimal. Conversely, if the strategy chooses not to issue a warning at node $1$, it violates the intended warning protocol, as the temperature reaches 0 in some samples.

This simple example demonstrates that a flawless warning system cannot be designed due to the probabilistic nature of the environment. Instead, the system should issue warnings when conditions indicate a high probability of need. \VI{In this case from the prefix $2;1$ indicated a decrease in temperature and the corresponding probability indicated a higher probability of reaching $0$  comparing to the other branches with prefix $2;2$, hence naturally suggested that an warning should be issued for prefix $2;1$ while not for prefix $2;2$. This distinction justifies our model with two types of constraints.}

\paragraph{Formalization}
We now turn to a formal description of the two types of constraints. The hard constraints are expressed using Linear Temporal Logic (LTL), while the soft constraints are defined with a reward machine (where negative rewards are interpreted as costs). This reward machine assigns negative rewards for false warnings (when the temperature does not drop to 0 after a warning) and imposes no penalty for accurate warnings. This balanced approach enables the system to optimize its responses while accounting for the inherent uncertainties in temperature evolution.

To formalize the scenario above, we begin by establishing notation and encoding conventions. First, we assume that the temperature is sampled frequently enough so that the temperature difference between any two consecutive inputs is at most $1$.
We focus only on the interval $[-1,2]$, with any value outside this interval abstracted as $-1$ or $2$, respectively. Furthermore, all values are rounded to the nearest integer to allow encoding as discrete inputs. Let $\{M_1, M_2\}$ be the set of input atomic propositions ${\sf AP}_I$; we assign temperatures to their binary encoding as follows:
{\begin{enumerate}
    \item $t=2\iff !M_1\,\&\,!M_2$
    \item $t=1\iff M_1\,\&\,!M_2$
    \item $t=0 \iff !M_1\,\&\,M_2$
    \item $t=-1 \iff M_1\,\&\,M_2$
\end{enumerate}
Let ${\sf AP}_O =\{ {\sf Warn}, {\sf Alarm} \}$ be the set of output atomic propositions, where the assignment of each atomic proposition corresponds to an action taken by the system. The case where both ${\sf Warn}$ and ${\sf Alarm}$ are active simultaneously is ruled out by encoding this restriction as part of the hard constraint $\varphi_{\sf hard}$.

The hard constraints are encoded as the conjunction of LTL formulas, each with its corresponding natural language interpretation:
\begin{enumerate}
    \item Alarm must be issued when the temperature is below or equal to $0$
    $$G(M_2 \rightarrow \sf {Alarm})$$
    \item Alarm can only be released when the temperature is above zero for two consecutive steps,
    $$G((!M_1 \,\&\, M_2) \,\&\, X(!M_2)) \rightarrow  X({\sf Alarm})$$
    $$G(!M_2 \,\&\, X(!M_2)) \rightarrow  X(!{\sf Alarm})$$
    \item Warning and Alarm should not be issued at the same time
    $$G(!{\sf Alarm}\,|\, !{\sf Warn})$$
\end{enumerate}

\begin{figure}[h!]
    \centering
    \begin{tikzpicture}[scale=0.15]
    \tikzstyle{every node}+=[inner sep=0pt]
    \draw [black] (18.3,-23.5) circle (3);
    \draw (18.3,-23.5) node {$s_0$};
    \draw [black] (39.7,-17.6) circle (3);
    \draw (39.7,-17.6) node {$s_2$};
    \draw [black] (50.1,-36) circle (3);
    \draw (50.1,-36) node {$s_1$};
    \draw [black] (26,-45.1) circle (3);
    \draw (26,-45.1) node {$s_3$};
    \draw [black] (36.999,-18.904) arc (-65.84125:-83.33159:53.594);
    \fill [black] (37,-18.9) -- (36.06,-18.77) -- (36.47,-19.69);
    \draw (26,-21.7) node [above] {\resizebox{0.06\textwidth}{!}{$2;\text{Warn},0$}};
    \draw [black] (21.277,-23.867) arc (81.59919:55.48304:62.757);
    \fill [black] (47.67,-34.24) -- (47.29,-33.38) -- (46.73,-34.2);
    \draw (32,-25.5) node [above] {\resizebox{0.06\textwidth}{!}{$1;\text{Warn},0$}};
    \draw [black] (42.545,-16.679) arc (99.82114:-40.86936:10.587);
    \fill [black] (52.36,-34.04) -- (53.26,-33.76) -- (52.5,-33.11);
    \draw (54.23,-20.69) node [right] {\resizebox{0.06\textwidth}{!}{$1;\text{Warn},0$}};
    \draw [black] (15.615,-22.188) arc (271.69424:-16.30576:2.25);
    \draw (11,-18) node [above] {\resizebox{0.06\textwidth}{!}{$2/1;\text{Off},0$}};
    \fill [black] (17.71,-20.57) -- (18.18,-19.76) -- (17.18,-19.79);
    \draw [black] (23.01,-45.197) arc (-95.51307:-225.24651:11.657);
    \fill [black] (23.01,-45.2) -- (22.26,-44.62) -- (22.17,-45.62);
    \draw (12.39,-38.28) node [left] {\resizebox{0.12\textwidth}{!}{$0/-1;\text{Alarm},-1$}};
    \draw [black] (40.322,-14.677) arc (195.70984:-92.29016:2.25);
    \draw (47,-12) node [above] {\resizebox{0.06\textwidth}{!}{$2;\text{Warn},0$}};
    \fill [black] (42.4,-16.32) -- (43.3,-16.58) -- (43.03,-15.62);
    \draw [black] (53.003,-35.29) arc (131.47119:-156.52881:2.25);
    \draw (57,-37.2) node [right] {\resizebox{0.06\textwidth}{!}{$1;\text{Warn},0$}};
    \fill [black] (52.43,-37.87) -- (52.58,-38.8) -- (53.33,-38.14);
    \draw [black] (48.041,-38.179) arc (-46.63475:-91.99284:26.413);
    \fill [black] (28.99,-45.37) -- (29.77,-45.9) -- (29.8,-44.9);
    \draw (43.77,-43.7) node [below] {\resizebox{0.06\textwidth}{!}{$0;\text{Alarm},0$}};
    \draw [black] (27.323,-47.78) arc (54:-234:2.25);
    \draw (26,-52.35) node [below] {\resizebox{0.12\textwidth}{!}{$1/0/-1;\text{Alarm},0$}};
    \fill [black] (24.68,-47.78) -- (23.8,-48.13) -- (24.61,-48.72);
    \draw [black] (24.99,-42.27) -- (19.31,-26.33);
    \fill [black] (19.31,-26.33) -- (19.11,-27.25) -- (20.05,-26.91);
    \draw (21.39,-35.07) node [left] {\resizebox{0.06\textwidth}{!}{$2/1;\text{Off},0$}};
    \draw [black] (39.14,-20.547) arc (-12.43977:-40.52354:51.413);
    \fill [black] (39.14,-20.55) -- (38.48,-21.22) -- (39.46,-21.44);
    \draw (32.5,-36) node [left] {\resizebox{0.06\textwidth}{!}{$2;\text{Warn},0$}};
    \draw [black] (28.626,-43.649) arc (117.84217:103.53024:79.565);
    \fill [black] (47.17,-36.65) -- (46.28,-36.35) -- (46.51,-37.32);
    \draw (40,-37.6) node [above] {\resizebox{0.06\textwidth}{!}{$1;\text{Warn},0$}};
    \draw [black] (20.659,-21.649) arc (124.71033:86.11683:25.217);
    \fill [black] (20.66,-21.65) -- (21.6,-21.6) -- (21.03,-20.78);
    \draw (24.58,-18) node [above] {\resizebox{0.07\textwidth}{!}{$2/1;\text{Off},-1$}};
    \draw [black] (47.175,-35.332) arc (-103.69491:-119.22285:104.506);
    \fill [black] (20.9,-25) -- (21.35,-25.83) -- (21.84,-24.96);
    \draw (29.16,-31) node [below] {\resizebox{0.07\textwidth}{!}{$2/1;\text{Off},-1$}};
    \draw [black] (48.62,-33.39) -- (41.18,-20.21);
    \fill [black] (41.18,-20.21) -- (41.13,-21.15) -- (42.01,-20.66);
    \draw (45,-25.58) node [right] {\resizebox{0.065\textwidth}{!}{$2;\text{Warn},-1$}};
\end{tikzpicture}
    \label{fig:RewM}
    \caption{The reward machine is designed to penalize false warning signals in hindsight.}
\end{figure}
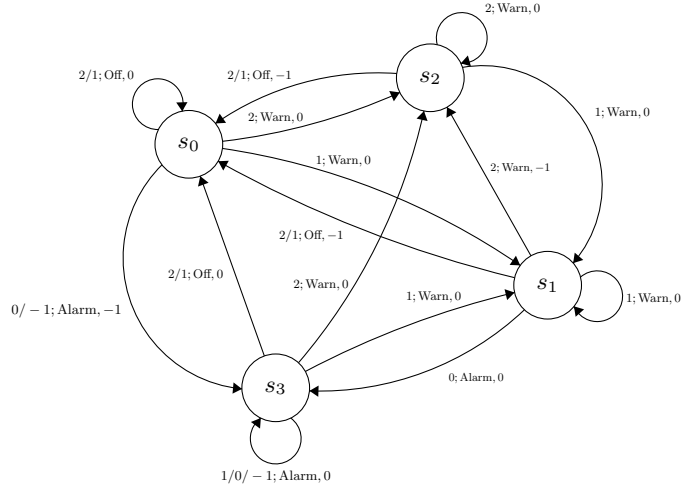\noindent

For the soft constraint on warnings, we construct the reward machine depicted in Fig.~\ref{fig:RewM}, where $s_0$ is the initial state, corresponding to no signal being issued; $s_1$ and $s_2$ correspond to states reached when a warning has been issued and the temperature is equal to $1$ and $2$, respectively; and $s_3$ corresponds to the state when an alarm has been issued. Intuitively, the following three types of behavior incur a reward of $-1$:
\begin{enumerate}
    \item {\sf Alarm} is directly issued without warning ahead, which corresponds to the edge from $s_0$ to $s_3$.
    \item {\sf Warn} is taken off without being reached to zero, which corresponds to the edge from $s_1$ and $s_2$ to $s_0$.
    \item The temperature increase while the {\sf Warn} is being issued, which corresponds to the arrow from $s_1$ to $s_0$.
\end{enumerate}

For readability, the edge labels are shown as temperature values and issued signals rather than their binary representations. Some input-output pairs are omitted here, as they correspond to actions that would violate the hard constraint $\varphi_{\sf hard}$. Our synthesis method will always avoid these edges and can therefore be safely omitted. 

\paragraph{Prototypal implementation} To demonstrate the practicality of our framework, we developed a prototype in Python. We used {\sc Acacia-Bonsai}~\cite{DBLP:conf/tacas/CadilhacP23} to translate the hard constraints expressed in LTL into universal coB\"uchi automata and employed its antichain solver to compute $\ceil{{\sf Win}}$. For constructing and solving the SMT instances, we utilized the {\sc PySMT} library~\cite{DBLP:conf/popl/LiAKGC14} with Z3~\cite{DBLP:conf/tacas/MouraB08} as the underlying solver. This setup allowed us to compute the optimal partial strategy on the sample tree, generating complete examples for a given set of samples. Finally, we applied the {\sf SynthLearn} algorithm~\cite{DBLP:conf/tacas/BalachanderFR23} to obtain a Mealy machine that realizes the hard constraints and generalizes the examples computed on the sample tree.

\begin{figure}[h!]
    \centering
    \begin{tikzpicture}[scale=0.1]
    \tikzstyle{every node}+=[inner sep=0pt]
    \draw [black] (15.5,-42.4) circle (3);
    \draw (15.5,-42.4) node {\resizebox{0.03\textwidth}{!}{$(-2,-1)$}};
    \draw [black] (15.5,-54.7) circle (3);
    \draw (15.5,-54.7) node {\resizebox{0.03\textwidth}{!}{$(-2,0)$}};
    \draw [black] (28.6,-54.7) circle (3);
    \draw (28.6,-54.7) node {\resizebox{0.03\textwidth}{!}{$(-2,1)$}};
    \draw [black] (15.5,-20.5) circle (3);
    \draw (15.5,-20.5) node {\resizebox{0.03\textwidth}{!}{$(-1,1)$}};
    \draw [black] (15.5,-7.9) circle (3);
    \draw (15.5,-7.9) node {\resizebox{0.03\textwidth}{!}{$(-1,0)$}};
    \draw [black] (28.6,-7.9) circle (3);
    \draw (28.6,-7.9) node {\resizebox{0.03\textwidth}{!}{$(-1,-1)$}};
    \draw [black] (32.6,-27.6) circle (3);
    \draw (32.6,-27.6) node {\resizebox{0.03\textwidth}{!}{$(0,1)$}};
    \draw [black] (40.9,-34.5) circle (3);
    \draw (40.9,-34.5) node {\resizebox{0.03\textwidth}{!}{$(0,0)$}};
    \draw [black] (47.8,-27.6) circle (3);
    \draw (47.8,-27.6) node {\resizebox{0.03\textwidth}{!}{$(0,-1)$}};
    \draw [black] (57.2,-7.9) circle (3);
    \draw (57.2,-7.9) node {\resizebox{0.03\textwidth}{!}{$(1,1)$}};
    \draw [black] (70.3,-20.5) circle (3);
    \draw (70.3,-20.5) node {\resizebox{0.03\textwidth}{!}{$(1,-1)$}};
    \draw [black] (70.3,-7.9) circle (3);
    \draw (70.3,-7.9) node {\resizebox{0.03\textwidth}{!}{$(1,0)$}};
    \draw [black] (70.3,-42.4) circle (3);
    \draw (70.3,-42.4) node {\resizebox{0.03\textwidth}{!}{$(2,1)$}};
    \draw [black] (70.3,-54.7) circle (3);
    \draw (70.3,-54.7) node {\resizebox{0.03\textwidth}{!}{$(2,0)$}};
    \draw [black] (57.2,-54.7) circle (3);
    \draw (57.2,-54.7) node {\resizebox{0.03\textwidth}{!}{$(2,-1)$}};
    \draw [blue] (16.512,-45.218) arc (13.65538:-13.65538:14.113);
    \fill [blue] (16.51,-51.88) -- (17.19,-51.22) -- (16.21,-50.99);
    
    \draw [blue] (25.6,-54.7) -- (18.5,-54.7);
    \fill [blue] (18.5,-54.7) -- (19.3,-55.2) -- (19.3,-54.2);
    
    \draw [green] (12.82,-43.723) arc (-36:-324:2.25);
    \fill [green] (12.82,-41.08) -- (12.47,-40.2) -- (11.88,-41.01);
    
    \draw [green] (27.36,-51.968) arc (-155.88595:-162.19628:278.288);
    \fill [green] (16.4,-23.36) -- (16.17,-24.28) -- (17.12,-23.97);
    
    \draw [red] (14.418,-39.603) arc (-162.09549:-197.90451:26.521);
    \fill [red] (14.42,-23.3) -- (13.7,-23.9) -- (14.65,-24.21);
    
    \draw [black] (13.862,-52.203) arc (-156.21491:-203.78509:9.058);
    \fill [black] (13.86,-44.9) -- (13.08,-45.43) -- (14,-45.83);
    
    \draw [black] (12.892,-53.223) arc (-124.08203:-235.91797:18.863);
    \fill [black] (12.89,-21.98) -- (11.95,-22.01) -- (12.51,-22.84);
    
    \draw [black] (14.131,-57.356) arc (0.45824:-287.54176:2.25);
    \fill [black] (12.56,-55.23) -- (11.75,-54.74) -- (11.76,-55.74);
    
    \draw [blue] (15.5,-17.5) -- (15.5,-10.9);
    \fill [blue] (15.5,-10.9) -- (15,-11.7) -- (16,-11.7);
    
    \draw [blue] (25.6,-7.9) -- (18.5,-7.9);
    \fill [blue] (18.5,-7.9) -- (19.3,-8.4) -- (19.3,-7.4);
    
    \draw [black] (13.332,-5.844) arc (254.24846:-33.75154:2.25);
    \fill [black] (15.81,-4.93) -- (16.51,-4.29) -- (15.55,-4.02);
    
    \draw [green] (16.185,-39.479) arc (166.13264:152.2828:128.376);
    \fill [green] (16.18,-39.48) -- (16.86,-38.82) -- (15.89,-38.58);
    
    \draw [red] (15.5,-23.5) -- (15.5,-39.4);
    \fill [red] (15.5,-39.4) -- (16,-38.6) -- (15,-38.6);
    
    \draw [black] (13.596,-40.084) arc (-143.9689:-216.0311:25.388);
    \fill [black] (13.6,-40.08) -- (13.53,-39.14) -- (12.72,-39.73);
    
    \draw [red] (31.006,-25.061) arc (-151.37112:-185.67362:24.57);
    \fill [red] (31.01,-25.06) -- (31.06,-24.12) -- (30.18,-24.6);
    
    \draw [green] (18.27,-21.65) -- (29.83,-26.45);
    \fill [green] (29.83,-26.45) -- (29.28,-25.68) -- (28.9,-26.6);
    
    \draw [black] (30.415,-25.544) arc (-134.01098:-144.07173:114.714);
    \fill [black] (30.42,-25.54) -- (30.19,-24.63) -- (29.49,-25.35);
    
    \draw [red] (26.41,-52.65) -- (17.69,-44.45);
    \fill [red] (17.69,-44.45) -- (17.93,-45.37) -- (18.61,-44.64);
    
    \draw [red] (32,-24.66) -- (29.2,-10.84);
    \fill [red] (29.2,-10.84) -- (28.87,-11.72) -- (29.85,-11.52);
    
    \draw [black] (30.946,-9.768) arc (48.58785:1.04436:29.703);
    \fill [black] (30.95,-9.77) -- (31.22,-10.67) -- (31.88,-9.92);
    
    \draw [green] (31.282,-9.243) arc (61.11375:27.41343:37.675);
    \fill [green] (31.28,-9.24) -- (31.74,-10.07) -- (32.22,-9.19);
    
    \draw [blue] (34.91,-29.52) -- (38.59,-32.58);
    \fill [blue] (38.59,-32.58) -- (38.3,-31.69) -- (37.66,-32.46);
    
    \draw [blue] (45.68,-29.72) -- (43.02,-32.38);
    \fill [blue] (43.02,-32.38) -- (43.94,-32.17) -- (43.23,-31.46);
    
    \draw [black] (42.223,-37.18) arc (54:-234:2.25);
    \fill [black] (39.58,-37.18) -- (38.7,-37.53) -- (39.51,-38.12);
    
    \draw [black] (40.931,-31.502) arc (-183.67005:-239.32838:27.902);
    \fill [black] (54.54,-9.29) -- (53.6,-9.27) -- (54.11,-10.13);
    
    \draw [red] (47.588,-24.61) arc (-180.33857:-230.67841:19.532);
    \fill [red] (54.74,-9.62) -- (53.81,-9.74) -- (54.44,-10.51);
    
    \draw [green] (33.328,-24.692) arc (162.38928:94.98712:24.1);
    \fill [green] (54.2,-7.97) -- (53.36,-7.55) -- (53.45,-8.54);
    
    \draw [green] (68.059,-22.492) arc (-51.72232:-93.25092:25.557);
    \fill [green] (50.78,-27.95) -- (51.55,-28.49) -- (51.61,-27.49);
    
    \draw [red] (57.029,-10.893) arc (-6.67081:-44.34616:25.206);
    \fill [red] (50.02,-25.58) -- (50.94,-25.36) -- (50.22,-24.66);
    
    \draw [black] (69.285,-10.722) arc (-22.83679:-74.75532:28.169);
    \fill [black] (50.73,-26.97) -- (51.63,-27.24) -- (51.37,-26.27);
    
    \draw [blue] (60.2,-7.9) -- (67.3,-7.9);
    \fill [blue] (67.3,-7.9) -- (66.5,-7.4) -- (66.5,-8.4);
    
    \draw [blue] (70.3,-17.5) -- (70.3,-10.9);
    \fill [blue] (70.3,-10.9) -- (69.8,-11.7) -- (70.8,-11.7);
    
    \draw [black] (70.252,-4.912) arc (208.65382:-79.34618:2.25);
    \fill [black] (72.64,-6.04) -- (73.58,-6.1) -- (73.1,-5.22);
    
    \draw [red] (72.093,-22.9) arc (31.50815:-31.50815:16.36);
    \fill [red] (72.09,-40) -- (72.94,-39.58) -- (72.08,-39.06);
    
    \draw [black] (72.398,-10.042) arc (40.69854:-40.69854:23.17);
    \fill [black] (72.4,-40.26) -- (73.3,-39.98) -- (72.54,-39.33);
    
    \draw [green] (58.26,-10.7) -- (69.24,-39.6);
    \fill [green] (69.24,-39.6) -- (69.42,-38.67) -- (68.48,-39.02);
    
    \draw [green] (57.157,-51.701) arc (-181.07097:-220.8468:45.64);
    \fill [green] (68.26,-22.7) -- (67.36,-22.98) -- (68.12,-23.64);
    
    \draw [red] (70.3,-39.4) -- (70.3,-23.5);
    \fill [red] (70.3,-23.5) -- (69.8,-24.3) -- (70.8,-24.3);
    
    \draw [black] (68.492,-52.308) arc (-146.16233:-213.83767:26.414);
    \fill [black] (68.49,-22.89) -- (67.63,-23.28) -- (68.46,-23.83);
    
    \draw [blue] (71.239,-45.244) arc (12.60062:-12.60062:15.154);
    \fill [blue] (71.24,-51.86) -- (71.9,-51.18) -- (70.93,-50.97);
    
    \draw [green] (73.101,-41.358) arc (138.14849:-149.85151:2.25);
    \fill [green] (72.83,-43.99) -- (73.09,-44.9) -- (73.76,-44.15);
    
    \draw [blue] (60.2,-54.7) -- (67.3,-54.7);
    \fill [blue] (67.3,-54.7) -- (66.5,-54.2) -- (66.5,-55.2);
    
    \draw [black] (73.254,-54.246) arc (126.47443:-161.52557:2.25);
    \fill [black] (72.46,-56.77) -- (72.53,-57.71) -- (73.33,-57.11);
    
    \draw [black] (68.826,-52.1) arc (-159.09755:-200.90245:9.95);
    \fill [black] (68.83,-45) -- (68.07,-45.57) -- (69.01,-45.93);
    
    \draw [red] (57.572,-51.731) arc (165.91089:100.48112:12.362);
    \fill [red] (67.31,-42.58) -- (66.44,-42.24) -- (66.62,-43.22);
    
\end{tikzpicture}
    \label{fig:Markov}
    \caption{The Markov Chain for modelling the environment}
\end{figure}
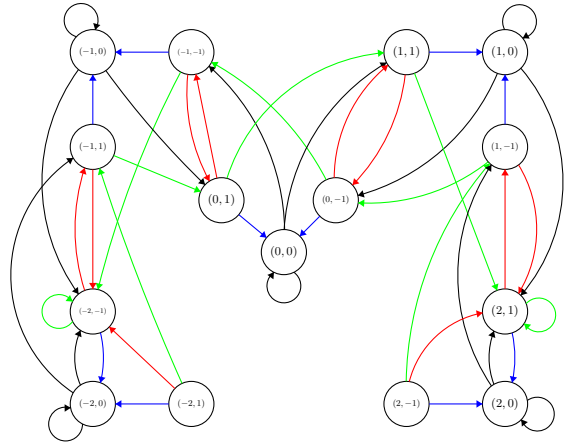\noindent

\begin{figure}[h!]
    \centering
    \includegraphics[width=0.9\textwidth]{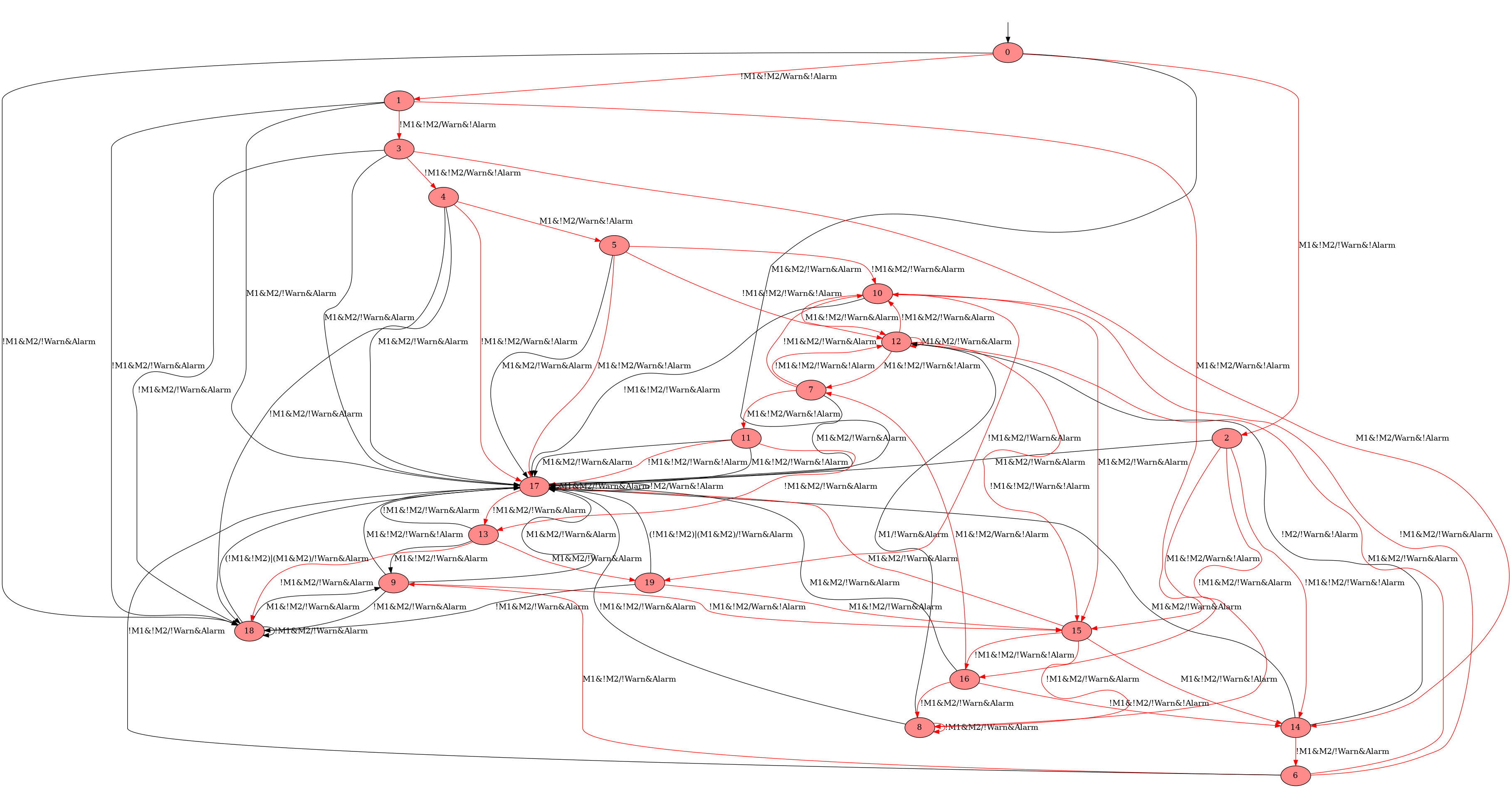}
    \label{fig:CMM}
    \caption{The completed Mealy machine}
\end{figure}\noindent

To enable the sampling of a representative environment, we modeled the environment as the finite Markov chain shown in Figure~\ref{fig:Markov}.The probability of each transition is defined according to the color of the edge, where $p_{\sf red}=p_{\sf blue}=\frac{1}{8}, p_{\sf green}=\frac{3}{4}, p_{\sf black}=\frac{1}{3}$. This model allowed us to draw representative samples. We used the parameters $n=100$ and $L=6$ for sampling. Our implementation successfully built a partial strategy on the sample tree constructed from these 100 samples, achieving an optimal value of $C = -0.25$ while maintaining the realizability of $\varphi_{\sf hard}$. When this partial function is applied to the 100 samples, it generates 100 complete examples, which are then passed to the {\sf SynthLearn} algorithm along with $\varphi_{\sf hard}$. This produces a complete Mealy machine that realizes $\varphi_{\sf hard}$ and generalizes the examples, resulting in a Mealy machine with 20 states, as depicted in Figure~\ref{fig:CMM}.

To evaluate the quality of the solution produced by {\sf SynthLearn}, we model-checked this machine within the environment shown in Fig.\ref{fig:Markov} and the reward machine in Fig.\ref{fig:RewM} using {\sc Storm} to compute the expected mean-payoff of the solution in the long run. The result was a long-run average reward of $-0.0707$, which is reasonably close to the sample tree's average reward of $\frac{-0.25}{6} = -0.0417$. This indicates that the {\sf SynthLearn} algorithm has effectively generalized the optimal local decisions computed from the samples to synthesize a complete strategy that enforces $\varphi_{\sf hard}$ with certainty while achieving strong performance in the environment. This outcome is particularly notable since we did not use the full model of the environment during synthesis, relying instead solely on samples drawn from it.

\section{Conclusion}
In this paper, we presented a framework for the automatic synthesis of reactive systems that must enforce {\em both} hard and soft constraints in an unknown, oblivious stochastic environment that can only be sampled. Our approach relies on an LTL specification to formalize the hard constraints, while a reward machine assigns rewards to finite sequences of inputs and outputs to capture the soft constraints.

Given a set of sampled input sequences from the environment, we demonstrated how to compute a partial strategy that is optimal with respect to the reward machine while preserving the realizability of the hard constraints. We established that computing this optimal strategy is computationally hard (the associated decision problem was shown {\sf NP-complete}), and provided an SMT-based solution to tackle this problem.

Once this partial strategy is applied to the sample, it produces complete examples, which can then be generalized using the {\sf SynthLearn} algorithm~\cite{DBLP:conf/tacas/BalachanderFR23}. This results in a complete strategy that not only enforces the hard constraints but also generalizes the behaviors illustrated by the examples.

We implemented a prototype tool to validate our framework and applied it to a simple yet insightful case study, showcasing its practical applicability and effectiveness.

\bibliographystyle{alpha}
\bibliography{biblio}

\newcommand{\etalchar}[1]{$^{#1}$}
\begin{thebibliography}{LAK{\etalchar{+}}14}

\bibitem[BBF{\etalchar{+}}12]{DBLP:conf/cav/BohyBFJR12}
Aaron Bohy, V{\'{e}}ronique Bruy{\`{e}}re, Emmanuel Filiot, Naiyong Jin, and Jean{-}Fran{\c{c}}ois Raskin.
\newblock Acacia+, a tool for {LTL} synthesis.
\newblock In P.~Madhusudan and Sanjit~A. Seshia, editors, {\em Computer Aided Verification - 24th International Conference, {CAV} 2012, Berkeley, CA, USA, July 7-13, 2012 Proceedings}, volume 7358 of {\em Lecture Notes in Computer Science}, pages 652--657. Springer, 2012.

\bibitem[BFR23]{DBLP:conf/tacas/BalachanderFR23}
Mrudula Balachander, Emmanuel Filiot, and Jean{-}Fran{\c{c}}ois Raskin.
\newblock {LTL} reactive synthesis with a few hints.
\newblock In {\em {TACAS} {(2)}}, volume 13994 of {\em Lecture Notes in Computer Science}, pages 309--328. Springer, 2023.

\bibitem[BFRR14]{DBLP:conf/stacs/BruyereFRR14}
V{\'{e}}ronique Bruy{\`{e}}re, Emmanuel Filiot, Mickael Randour, and Jean{-}Fran{\c{c}}ois Raskin.
\newblock Meet your expectations with guarantees: Beyond worst-case synthesis in quantitative games.
\newblock In Ernst~W. Mayr and Natacha Portier, editors, {\em 31st International Symposium on Theoretical Aspects of Computer Science {(STACS} 2014), {STACS} 2014, March 5-8, 2014, Lyon, France}, volume~25 of {\em LIPIcs}, pages 199--213. Schloss Dagstuhl - Leibniz-Zentrum f{\"{u}}r Informatik, 2014.

\bibitem[BK08]{DBLP:books/daglib/0020348}
Christel Baier and Joost{-}Pieter Katoen.
\newblock {\em Principles of model checking}.
\newblock {MIT} Press, 2008.

\bibitem[CP23]{DBLP:conf/tacas/CadilhacP23}
Micha{\"{e}}l Cadilhac and Guillermo~A. P{\'{e}}rez.
\newblock Acacia-bonsai: {A} modern implementation of downset-based {LTL} realizability.
\newblock In Sriram Sankaranarayanan and Natasha Sharygina, editors, {\em Tools and Algorithms for the Construction and Analysis of Systems - 29th International Conference, {TACAS} 2023, Held as Part of the European Joint Conferences on Theory and Practice of Software, {ETAPS} 2022, Paris, France, April 22-27, 2023, Proceedings, Part {II}}, volume 13994 of {\em Lecture Notes in Computer Science}, pages 192--207. Springer, 2023.

\bibitem[CRR14]{DBLP:journals/acta/ChatterjeeRR14}
Krishnendu Chatterjee, Mickael Randour, and Jean{-}Fran{\c{c}}ois Raskin.
\newblock Strategy synthesis for multi-dimensional quantitative objectives.
\newblock {\em Acta Informatica}, 51(3-4):129--163, 2014.

\bibitem[dMB08]{DBLP:conf/tacas/MouraB08}
Leonardo~Mendon{\c{c}}a de~Moura and Nikolaj~S. Bj{\o}rner.
\newblock {Z3:} an efficient {SMT} solver.
\newblock In {\em {TACAS}}, volume 4963 of {\em Lecture Notes in Computer Science}, pages 337--340. Springer, 2008.

\bibitem[FFT18]{DBLP:journals/corr/abs-1803-09566}
Peter Faymonville, Bernd Finkbeiner, and Leander Tentrup.
\newblock Bosy: An experimentation framework for bounded synthesis.
\newblock {\em CoRR}, abs/1803.09566, 2018.

\bibitem[FJR09]{DBLP:conf/cav/FiliotJR09}
Emmanuel Filiot, Naiyong Jin, and Jean{-}Fran{\c{c}}ois Raskin.
\newblock An antichain algorithm for {LTL} realizability.
\newblock In Ahmed Bouajjani and Oded Maler, editors, {\em Computer Aided Verification, 21st International Conference, {CAV} 2009, Grenoble, France, June 26 - July 2, 2009. Proceedings}, volume 5643 of {\em Lecture Notes in Computer Science}, pages 263--277. Springer, 2009.

\bibitem[FJR11]{DBLP:journals/fmsd/FiliotJR11}
Emmanuel Filiot, Naiyong Jin, and Jean{-}Fran{\c{c}}ois Raskin.
\newblock Antichains and compositional algorithms for {LTL} synthesis.
\newblock {\em Formal Methods Syst. Des.}, 39(3):261--296, 2011.

\bibitem[HJK{\etalchar{+}}22]{DBLP:journals/sttt/HenselJKQV22}
Christian Hensel, Sebastian Junges, Joost{-}Pieter Katoen, Tim Quatmann, and Matthias Volk.
\newblock The probabilistic model checker storm.
\newblock {\em Int. J. Softw. Tools Technol. Transf.}, 24(4):589--610, 2022.

\bibitem[KNP09]{DBLP:journals/sigmetrics/KwiatkowskaNP09}
Marta~Z. Kwiatkowska, Gethin Norman, and David Parker.
\newblock {PRISM:} probabilistic model checking for performance and reliability analysis.
\newblock {\em {SIGMETRICS} Perform. Evaluation Rev.}, 36(4):40--45, 2009.

\bibitem[KV05]{DBLP:conf/focs/KupfermanV05}
Orna Kupferman and Moshe~Y. Vardi.
\newblock Safraless decision procedures.
\newblock In {\em 46th Annual {IEEE} Symposium on Foundations of Computer Science {(FOCS} 2005), 23-25 October 2005, Pittsburgh, PA, USA, Proceedings}, pages 531--542. {IEEE} Computer Society, 2005.

\bibitem[LAK{\etalchar{+}}14]{DBLP:conf/popl/LiAKGC14}
Yi~Li, Aws Albarghouthi, Zachary Kincaid, Arie Gurfinkel, and Marsha Chechik.
\newblock Symbolic optimization with {SMT} solvers.
\newblock In {\em {POPL}}, pages 607--618. {ACM}, 2014.

\bibitem[LMS20]{DBLP:journals/acta/LuttenbergerMS20}
Michael Luttenberger, Philipp~J. Meyer, and Salomon Sickert.
\newblock Practical synthesis of reactive systems from {LTL} specifications via parity games.
\newblock {\em Acta Informatica}, 57(1-2):3--36, 2020.

\bibitem[PR89]{DBLP:conf/popl/PnueliR89}
Amir Pnueli and Roni Rosner.
\newblock On the synthesis of a reactive module.
\newblock In {\em Conference Record of the Sixteenth Annual {ACM} Symposium on Principles of Programming Languages, Austin, Texas, USA, January 11-13, 1989}, pages 179--190. {ACM} Press, 1989.

\bibitem[QK21]{DBLP:conf/tacas/QuatmannK21}
Tim Quatmann and Joost{-}Pieter Katoen.
\newblock Multi-objective optimization of long-run average and total rewards.
\newblock In Jan~Friso Groote and Kim~Guldstrand Larsen, editors, {\em Tools and Algorithms for the Construction and Analysis of Systems - 27th International Conference, {TACAS} 2021, Held as Part of the European Joint Conferences on Theory and Practice of Software, {ETAPS} 2021, Luxembourg City, Luxembourg, March 27 - April 1, 2021, Proceedings, Part {I}}, volume 12651 of {\em Lecture Notes in Computer Science}, pages 230--249. Springer, 2021.

\bibitem[RRS17]{DBLP:journals/fmsd/RandourRS17}
Mickael Randour, Jean{-}Fran{\c{c}}ois Raskin, and Ocan Sankur.
\newblock Percentile queries in multi-dimensional markov decision processes.
\newblock {\em Formal Methods Syst. Des.}, 50(2-3):207--248, 2017.

\bibitem[SF07]{DBLP:conf/atva/ScheweF07a}
Sven Schewe and Bernd Finkbeiner.
\newblock Bounded synthesis.
\newblock In Kedar~S. Namjoshi, Tomohiro Yoneda, Teruo Higashino, and Yoshio Okamura, editors, {\em Automated Technology for Verification and Analysis, 5th International Symposium, {ATVA} 2007, Tokyo, Japan, October 22-25, 2007, Proceedings}, volume 4762 of {\em Lecture Notes in Computer Science}, pages 474--488. Springer, 2007.

\end{thebibliography}

\end{document}